\documentclass[linenumbers,twocolumn]{aastex631}

\newcommand{\N}[1]{\textcolor{blue}{ #1}} 

\usepackage{graphicx}
\usepackage{tablefootnote}
\usepackage{txfonts}

\usepackage{hyperref}
\hypersetup{
    colorlinks=true,
    linkcolor=blue,
    filecolor=magenta,      
    urlcolor=cyan,
    pdftitle={Overleaf Example},
    pdfpagemode=FullScreen,
    }

\begin{document}

\title{The Green Monster hiding in front of Cas\,A: \\\textit{JWST} reveals a dense and dusty circumstellar structure pockmarked by ejecta interactions}

\author[0000-0001-9419-6355]{Ilse De Looze}
\affiliation{Sterrenkundig Observatorium, Ghent University, Krijgslaan 281 - S9, B-9000 Gent, Belgium}

\author[0000-0002-0763-3885]{Dan Milisavljevic}
\affiliation{Purdue University, Department of Physics and Astronomy, 525 Northwestern Ave, West Lafayette, IN 47907, USA}
\affiliation{Integrative Data Science Initiative, Purdue University, West Lafayette, IN 47907, USA}

\author[0000-0001-7380-3144]{Tea Temim}
\affiliation{Princeton University, 4 Ivy Ln, Princeton, NJ 08544, USA}

\author[0000-0003-0913-4120]{Danielle Dickinson}
\affiliation{Purdue University, Department of Physics and Astronomy, 525 Northwestern Ave, West Lafayette, IN 47907 }

\author[0000-0003-3829-2056]{Robert Fesen}
\affiliation{6127 Wilder Lab, Department of Physics and Astronomy, Dartmouth College, Hanover, NH 03755, USA}

\author[0000-0001-8403-8548]{Richard G. Arendt}
\affiliation{Center for Space Sciences and Technology, University of Maryland, Baltimore County, Baltimore, MD 21250, USA}
\affiliation{Code 665, NASA/GSFC, 8800 Greenbelt Road, Greenbelt, MD 20771, USA}
\affiliation{Center for Research and Exploration in Space Science and Technology, NASA/GSFC, Greenbelt, MD 20771, USA}

\author[0000-0001-9419-6355]{Jeremy Chastenet}
\affiliation{Sterrenkundig Observatorium, Ghent University, Krijgslaan 281 - S9, B-9000 Gent, Belgium}

\author[0000-0003-2836-540X]{Salvatore Orlando}
\affiliation{INAF – Osservatorio Astronomico di Palermo, Piazza del Parlamento 1, 90134 Palermo, Italy}

\author[0000-0002-4708-4219]{Jacco Vink}
\affiliation{Anton Pannekoek Institute for Astronomy \& GRAPPA, University of Amsterdam, Science Park 904, 1098 XH Amsterdam, The Netherlands}
\affiliation{SRON Netherlands Institute for Space Research, Niels Bohrweg 4, 2333 CA Leiden, the Netherlands }

\author[0000-0002-3875-1171]{Michael J.\ Barlow}
\affiliation{Department of Physics and Astronomy, University College London, Gower Street, London WC1E 6BT, United Kingdom}

\author[0000-0002-3036-0184]{Florian Kirchschlager}
\affiliation{Sterrenkundig Observatorium, Ghent University, Krijgslaan 281 - S9, B-9000 Gent, Belgium}

\author{Felix D.\ Priestley}
\affiliation{Cardiff Hub for Astrophysical Research and Technology (CHART), School of Physics \& Astronomy, Cardiff University, The Parade, Cardiff CF24 3AA, UK}

\author[0000-0002-7868-1622]{John C.\ Raymond}
\affiliation{Center for Astrophysics $\vert$ Harvard \& Smithsonian, 60 Garden Street, Cambridge, MA 02138, USA}

\author[0000-0003-3643-839X]{Jeonghee Rho}
\affiliation{SETI Institute, 339 Bernardo Ave., Ste. 200, Mountain View, CA 94043, USA}
\affiliation{Department of Physics and Astronomy, Seoul National University, Gwanak-ro 1, Gwanak-gu, Seoul, 08826, South Korea}

\author[0000-0003-2138-5192]{Nina S. Sartorio}
\affiliation{Sterrenkundig Observatorium, Ghent University, Krijgslaan 281 - S9, B-9000 Gent, Belgium}

\author[0009-0002-3600-4516]{Tassilo Scheffler}
\affiliation{Sterrenkundig Observatorium, Ghent University, Krijgslaan 281 - S9, B-9000 Gent, Belgium}

\author{Franziska Schmidt}
\affiliation{Department of Physics and Astronomy, University College London, Gower Street, London WC1E 6BT, United Kingdom}

\author[0000-0003-2379-6518]{William P.\ Blair}
\affiliation{The William H. Miller III Department of Physics and Astronomy, Johns Hopkins University, 3400 N. Charles Street, Baltimore, MD 21218, USA}

\author[0000-0003-2238-1572]{Ori Fox}
\affiliation{Space Telescope Science Institute, 3700 San Martin Drive, Baltimore, MD 21218, USA}

\author{Christopher Fryer}
\affiliation{Center for Theoretical Astrophysics, Los Alamos National Laboratory, Los Alamos, NM 87545, USA}
\affiliation{Department of Astronomy, The University of Arizona, Tucson, AZ 85721, USA}
\affiliation{Department of Physics and Astronomy, The University of New Mexico, Albuquerque, NM 87131, USA}
\affiliation{Department of Physics, The George Washington University, Washington, DC 20052, USA}

\author[0000-0002-0831-3330]{Hans-Thomas Janka}
\affiliation{Max-Planck-Institut für Astrophysik, Karl-Schwarzschild-Str. 1, 85748, Garching, Germany}

\author[0000-0002-2755-1879]{Bon-Chul Koo}
\affiliation{Department of Physics and Astronomy, Seoul National University, Seoul 08861, Republic of Korea}

\author[0000-0002-3362-7040]{J.\ Martin Laming}
\affiliation{Space Science Division, Code 7684, Naval Research Laboratory, Washington, DC 20375, USA}

\author[0000-0002-5529-5593]{Mikako Matsuura}
\affiliation{Cardiff Hub for Astrophysical Research and Technology (CHART), School of Physics \& Astronomy, Cardiff University, The Parade, Cardiff CF24 3AA, UK}

\author[0000-0002-7507-8115]{Dan Patnaude}
\affiliation{Center for Astrophysics $\vert$ Harvard \& Smithsonian, 60 Garden Street, Cambridge, MA 02138, USA}

\author[0000-0003-1682-1148]{M\'onica Rela\~no}
\affiliation{Dept. F\'isica Te\'orica y del Cosmos, Universidad de Granada, 18071, Granada, Spain }

\newcommand{\STSCI}{\affiliation{Space Telescope Science Institute, Baltimore, MD 21218, USA}}
\newcommand{\JHU}{\affiliation{Department of Physics and Astronomy, The Johns Hopkins University, Baltimore, MD 21218, USA}}
\author[0000-0002-4410-5387]{A.~Rest}
\STSCI \JHU

\author[0000-0002-2617-5517]{Judy Schmidt}
\affiliation{Astrophysics Source Code Library, Michigan Technological University, 1400 Townsend Drive, Houghton, MI 49931, USA }

\author[0000-0001-5510-2424]{Nathan Smith}
\affil{Steward Observatory, University of Arizona, 933 N. Cherry Avenue, Tucson, AZ 85721, USA}

\author{Niharika Sravan}
\affil{Department of Physics, Drexel University, Philadelphia, PA 19104, USA}



\begin{abstract}

\noindent
{\sl JWST} observations of the young Galactic supernova remnant Cassiopeia~A revealed an unexpected structure seen as a green emission feature in colored composite MIRI F1130W and F1280W images -- hence dubbed the Green Monster -- that stretches across the central parts of the remnant in projection. Combining the kinematic information from NIRSpec and MIRI MRS with the multi-wavelength imaging from NIRCam and MIRI, we associate the Green Monster with circumstellar material that was lost during an asymmetric mass-loss phase. MIRI images are dominated by dust emission but its spectra show emission lines from Ne, H and Fe with low radial velocities indicative of a CSM nature. An X-ray analysis of this feature in a companion paper \citep{Vink2024} supports its CSM nature and detects significant blue shifting, thereby placing the Green Monster on the near side, in front of the Cas~A SN remnant. 
The most striking features of the Green Monster are dozens of almost perfectly circular $1''-3\arcsec$ sized holes, most likely created by interaction between high-velocity SN ejecta material and the CSM. Further investigation is needed to understand whether these holes were formed by small 8000--10500 km s$^{-1}$ N-rich ejecta knots that penetrated and advanced out ahead of the remnant's 5000 - 6000 km s$^{-1}$ outer blastwave, or by narrow ejecta fingers that protrude into the forward-shocked CSM. The detection of the Green Monster provides further evidence of the highly asymmetric mass-loss that Cas\,A's progenitor star underwent prior to its explosion.

\end{abstract}



\section{Introduction} \label{sec:intro}

Cassiopeia A (Cas\,A) is one of the best studied core-collapse supernova remnants (SNR) in our Galaxy due to both it youth and relatively short distance of $\sim$3.4 kpc \citep{Reed1995,Alarie2014}. 
Detailed studies of Cas\,A using multi-wavelength imaging and spectroscopic campaigns have given us unprecedented insights into supernova explosion asymmetries \citep{Fesen2001,Hwang2004,Grefenstette2014,Milisavljevic2015}, dust formation in the supernova ejecta \citep{Rho2008,Barlow2010,DeLooze2017,Bevan2017,Priestley2022}, the subsequent destruction by the reverse shock \citep{Biscaro2016,Bocchio2016,Micelotta2016,Kirchschlager2019,Kirchschlager2023}, and particle acceleration in supernova forward and reverse shocks \citep{Helder2008}. 

While no firm records exist of the sighting of the explosion of Cas\,A nearly 350 years ago \citep{Thorstensen2001,Koo2017}, the post-explosion classification of Cas\,A as a Type IIb supernova based on light echo spectra \citep{Krause2008, Rest2008, Rest2011} suggests that the Cas\,A progenitor experienced significant mass loss, having shed most of its hydrogen layer prior to explosion. Such heavy mass loss from a massive progenitor star 
\citep[$15-25$\,M$_{\odot}$;][]{Fesen1987,Fesen2001,Young2006} has possibly occurred due to interactions with a binary companion \citep{Chevalier2003,Young2006,Krause2008,Weil2020}. 
However, deep searches have not yet resulted in the detection of a companion star \citep{Kochanek2018, Kerzendorf2019}, leaving the possibility that the binary star is a faint low mass dwarf or that the two stars merged prior to the supernova explosion (e.g., \citealt{Nomoto1993}).

Studying the mass loss history of Cas\,A --- with constraints on the timing, the rate and the asymmetry of the mass loss  --- can give important insights into the pre-explosion conditions. There is evidence in the literature for several mass loss phases. Diffuse emission clouds about $10-20$ pc to the north, east and northwest of the explosion center detected in optical data are thought to arise from a slow (10 km s$^{-1}$) red supergiant wind \citep{Weil2020}. Studies of the X-ray ejecta knots are consistent with Cas\,A expanding into a slow RSG wind \citep{Laming2003,Hwang2009} with a relatively high density ($\sim$1\,cm$^{-3}$) for the pre-shock wind material \citep{Lee2014} at the current outer radius ($\sim$3\,pc) of Cas\,A. 
Faint dust emission was detected with \textit{Spitzer} and \textit{Herschel} roughly 2 to 3 pc north and east of the explosion center \citep{Barlow2010,Arendt2014}, and corresponds to the infrared counterpart of the X-ray emission that was thought to originate from a tenuous slow RSG wind. 

With the advent of the James Webb Space Telescope (\textit{JWST}) we can now map out the details of CSM dust structures shocked by the supernova blastwave \citep{Milisavljevic2024}. The circumstellar dust emission detected north, east and west of Cas\,A stands out as diffuse red emission around the periphery of the shocked ejecta shell in the \textit{JWST} multi-color image (see Figure\,\ref{FigWebbJudy}). Although the coverage towards the south does not extend much beyond 2 pc, filamentary structures trace out CSM that seems to follow the onset of an arc-like structure populated with dense circumstellar clumps often referred to as quasi-stationary flocculi (see Figure\,\ref{FigQSFs}).

The nature of these quasi-stationary flocculi (QSFs), first identified in optical images 
 starting in the early 1950s \citep{Baade1954,Minkowski1959,vandenBergh1970,Peimbert1971,vandenBergh1971,Kamper1976}, is less clear. QSFs are generally considered to be dense, pre-SN CSM knots with relatively low radial velocities of between $-500$ and +100 km s$^{-1}$ \citep{vandenBergh1985,Alarie2014,Koo2018} 
which led to them being called ``quasi-stationary" compared to Cas~A's 
fast-moving knots (FMKs) of SN ejecta. 
Detailed studies of optical and near-infrared data suggest that these QSFs have a high density (10$^{3}$-10$^{4}$ cm$^{-3}$, \citealt{Koo2020}), requiring recent (forward) shock interaction to make them detectable at optical and NIR wavelengths.

The He and N enrichment of QSFs \citep{Chevalier1978,Lamb1978,Koo2023} is suggestive of CNO-processed CSM material originating from the H-burning shell of the progenitor, and must have been generated at an evolved stage of stellar evolution for elements near the \text{H-burning shell} to reach the surface, hence after a significant portion of the hydrogen envelope was stripped. 
Indeed, proper motions of several QSFs suggest that the material was lost by the progenitor star roughly 10$^{4}$ years prior to explosion \citep{Peimbert1971,Kamper1976,
Chevalier1978,vandenBergh1985}.
However, this value should be considered with caution given that recent  interaction with Cas~A's forward shock front may have influenced the QSFs' currently observed motion. 

It is unclear how these dense circumstellar clumps with unshocked densities of $\sim$1000 cm$^{-3}$ formed in the first place. 
In one scenario, the QSFs originate from hydrodynamical instabilities that occurred in a dense shell of the RSG wind that got compressed by a fast wind in the final blue/yellow supergiant or Wolf-Rayet phase of the progenitor prior to the explosion. The dense remainders of this RSG wind-driven shell could be the QSFs shocked by the supernova blast wave that we observe today. 

However, the QSFs are distributed asymmetrically around the remnant, and do not seem to follow a shell structure \citep{Alarie2014,Koo2018}. It has been suggested that QSFs could be dense clumps, formed as overdensities in a smooth RSG wind, rather than the fragments of a disrupted circumstellar shell \citep{Chevalier2003}. The QSFs are also reminiscent of the slow N-rich condensations around $\eta$ Carinae \citep{sm04}, raising the possibility that they may have resulted instead from episodic mass loss during binary interaction.

Studies of the X-ray properties of shocked outer ejecta knots suggest there may have been a brief post-RSG phase that created a small (0.2--0.3\,pc) low density cavity around Cas\,A \citep{Hwang2009}. The presence of the north-east jet constrains the duration of such a high-velocity wind to significantly less than 10,000 years \citep{Schure2008}. Such a low density cavity could have formed from a tenuous fast wind originating from a Wolf-Rayet star or a yellow supergiant \citep{Weil2020}. But a Wolf-Rayet star scenario may be considered less likely due to the difficulty to explain the formation of QSFs by a WR wind \citep{vanVeelen2009} and due to the presence of H in the light echo spectra suggesting that some H was left after the mass loss \citep{Krause2008, Rest2008, Rest2011}. In general, Wolf-Rayet stars tend to be associated with more massive ($>$40\,M$_{\odot}$) progenitor stars \citep{Humphreys1985}, which would be at odds with the currently predicted progenitor mass (15--25\,M$_{\odot}$) of Cas~A.   

The spatially varying velocities of the reverse shock across the remnant, and the backwards motion of the reverse shock towards the West 
\citep{Fesen2019,Vink2022} also suggest the forward shock initially traversed a lopsided cavity. However, the backward motion of the reverse shock could also be explained through the interaction of the SN blast wave with a partial dense circumstellar shell $\approx 200$~years after the explosion \citep{Vink2022,Orlando2022}. The presence of such a fast wind during a brief blue or yellow supergiant or WR phase has been suggested both from observations and simulations \citep{Hwang2009, Koo2020,Weil2020,Vink2022,Orlando2022}, but again the connection of this heavy mass loss episode with the presence of QSFs and other mass loss phases remains unclear.

\textit{JWST} \citep{Rigby2023,Gardner2023} is an ideal facility to map out the shocked and unshocked ejecta structures in Cas\,A \citep{Milisavljevic2024} and to study its pre-SN mass loss history of Cas\,A. The shock-heated circumstellar (CSM) warm dust emission lights up at mid-infrared wavelengths (see Figure \ref{FigWebbJudy}), and the [Fe\,{\sc{ii}}]\,1.644\,$\mu$m line emission probed with NIRCam allows to study both dense shocked CSM and shocked ejecta clumps (Koo et al.\,in prep). The \textit{JWST} data, furthermore, reveal the spatial distribution of CO molecules that reformed in the ejecta after reverse shock processing \citep{Rho2024} and provide unique insights into the chemical diversity of supernova dust formation (Temim et al.\,in prep).

Each new telescope facility observing Cas\,A made surprising discoveries, contributing to our understanding of supernova explosion physics and pre-SN mass loss. One of the most striking features of the new \textit{JWST}/NIRCam and MIRI observations of Cas\,A (see Figure \ref{FigWebbJudy}) is a prominent structure --- dubbed the ``Green Monster\footnote{Cas~A's Green Monster was named after the tall, imposing green wall in the left field of Fenway Park, home of the Boston Red Sox baseball team.}" (hereafter GM) due to its green color in the multi-color press release image of Cas A, shown in Figure 1. The emission from this region is particularly bright in the MIRI F1130W and F1280W filters and dominates the central region of the SNR. 
Apart from its color, the GM stands out for hosting peculiar circular structures. In this paper, we use MIRI MRS and NIRSpec observations of one specific targeted ring structure in the GM --- in addition to NIRCam and MIRI photometric mapping of the entire GM structure --- to study the nature of the mid-infrared (MIR) emitting material. 

The observational details are described in Section \ref{Obs.sec}. The observational characteristics of the GM as seen by the JWST are presented in Section \ref{GMobs.sec}. The nature of the GM emission is discussed in Section \ref{GM.sec}, along with several scenarios that can account for the creation of the holes and rings. We discuss how the GM's structure fits in the overall mass loss history of Cas\,A (Section \ref{CSM.sec}) before concluding (Section \ref{Concl.sec}). Appendices \ref{LineProfile.sec} and \ref{LineFit.sec} provide a detailed explanation on how the line velocity profiles and maps were obtained. 

\section{Observations}
\label{Obs.sec}
\subsection{JWST/MIRI and JWST/NIRCam imaging data}
The Galactic supernova remnant Cas\,A was observed with \textit{JWST}/MIRI \citep{Wright2023} and \textit{JWST}/NIRCam \citep{Rieke2023} during August 2022 and November 2022, covering the entire remnant and part of its surroundings, as part of the Cycle 1 program GO01947 (PI: D. Milisavljevic). A return visit was required for MIRI to cover gaps between the detector footprints, and took place in November 2022. A full description of the observational details and data reduction pipeline is given in \citet{Milisavljevic2024}. We briefly summarize the most important information here. We obtained data in a $3\times5$ (fields of view) mosaic for all MIRI filters --- except for the F1500W filter\footnote{We obtained partial coverage for the F1500W filter from simultaneous MIRI imaging during the MRS observations.} --- providing us with good coverage of the mid-infrared spectral energy distribution (SED). We used the FASTR1 readout mode, with eight groups and one integration per exposure and a 4-point dither pattern, amounting to a total exposure time of 88.8\,s per individual field. A dedicated background observation was obtained and the background emission was subtracted during level 2-pipeline processing (except for the F770W, F1000W, F1130W and F1280W filters which show highly variable background emission levels). Aside from standard data reduction procedures --- but skipping the use of the \texttt{tweakreg} and \texttt{skymatch} steps ---  the regular and return visits were combined into a single mosaic after applying corrections for astrometric offsets using GAIA DR3 catalogs based on the output from the \textit{JWST} HST Alignment Tool \citep[JHAT;][]{Rest2023}. These astrometric corrections were fine-tuned on the highest-resolution F560W image, and applied to the other MIRI filters.  Due to small-scale spatial variations in the mid-infrared interstellar dust emission in the background of Cas\,A \citep[e.g.,][]{DeLooze2017},
we required a separate background subtraction routine to match the zero-levels of all eight MIRI images. Note that we did not subtract the variable background emission which would require detailed modelling efforts to disentangle the different sources of emission (interstellar, circumstellar and ejecta line and dust emission). Instead, we selected regions with no obvious background emission outside of the remnant and scaled all the MIRI images to have zero flux within these regions.  

The \textit{JWST}/NIRCam observations were designed to study the [Fe\,{\sc{ii}}] 1.644\,$\mu$m line and the 4.5\,$\mu$m CO fundamental band using one medium (F162M) and two wide (F356W, F444W) filters, respectively. These observations are discussed and analysed in more detail in \citet{Rho2024} and Koo et al.\,(in prep). In brief, the remnant was covered with a $3\times1$ (fields of view) mosaic using the BRIGHT1 readout pattern with seven groups and one integration per exposure. The total exposure time was 1675\,s for a total of 12 dithers per field using the 3TIGHT primary dithers, including 4 subpixel dithers. The exposure time was doubled for the F162M filter since this Short Wavelength Camera filter was covered during both of the F356W and F444W observations with the Long Wavelength Camera. The image processing for NIRCam was done using the calibration pipeline (v\,1.8.4) with the calibration reference data system v\,11.16.14, in a similar way as was done for the MIRI imaging data.

The spatial resolution (expressed in terms of the full-width-half-maximum (FWHM) of the PSF) achieved with NIRCam ($0.055\,\arcsec$, $0.116\,\arcsec$ and $0.145\,\arcsec$ for the F162M, F356W and F444W filters) and MIRI (ranging from $0.207\,\arcsec$ for F560W to $0.803\,\arcsec$ for F2550W) corresponds to physical scales ranging between 190\,au ($2.8\times10^{15}$\,cm) and 2700\,au ($4.1\times10^{16}$\,cm) --- if we assume a distance to Cas\,A of 3.4\,kpc. NIRCam filters are dominated by synchrotron emission and line emission --- mostly from the [Fe\,{\sc{ii}}]\,1.644\,$\mu$m line and the CO fundamental band at 4.6\,$\mu$m, but fainter lines ([Si\,{\sc{i}}]\,1.645\,$\mu$m, [Si\,{\sc{ix}}]\,3.94\,$\mu$m and [Mg\,{\sc{iv}}]\,4.49\,$\mu$m) and emission from polycyclic aromatic hydrocarbons and hot carbonaceous dust can also have non-negligible contributions in certain regions. While the F560W image is largely dominated by synchrotron emission, all other MIRI filters have a stronger contribution from dust emission originating from different sources (circumstellar, interstellar, and ejecta dust). Various MIRI filters also have non-negligible contributions from line emission \citep[see Table 1 in][for an overview]{Milisavljevic2024}. Most of these lines (e.g., [Ar\,{\sc{ii}}] 6.99\,$\mu$m, [Ar\,{\sc{iii}}] 8.99\,$\mu$m, [S\,{\sc{iv}}] 10.51\,$\mu$m, [Ne\,{\sc{ii}}] 12.81\,$\mu$m, [S\,{\sc{iii}}] 18.71\,$\mu$m and [O\,{\sc{iv}}] 25.89\,$\mu$m) originate from the supernova ejecta, but some lines ([Ne\,{\sc{ii}}] 12.81\,$\mu$m, [Ne\,{\sc{iii}}] 15.56\,$\mu$m, H$_{2}$ (0,0) S(1) 17.03\,$\mu$m) may have a circum- or interstellar origin. 

\begin{figure*}
\centering
\includegraphics[width=0.95\textwidth]{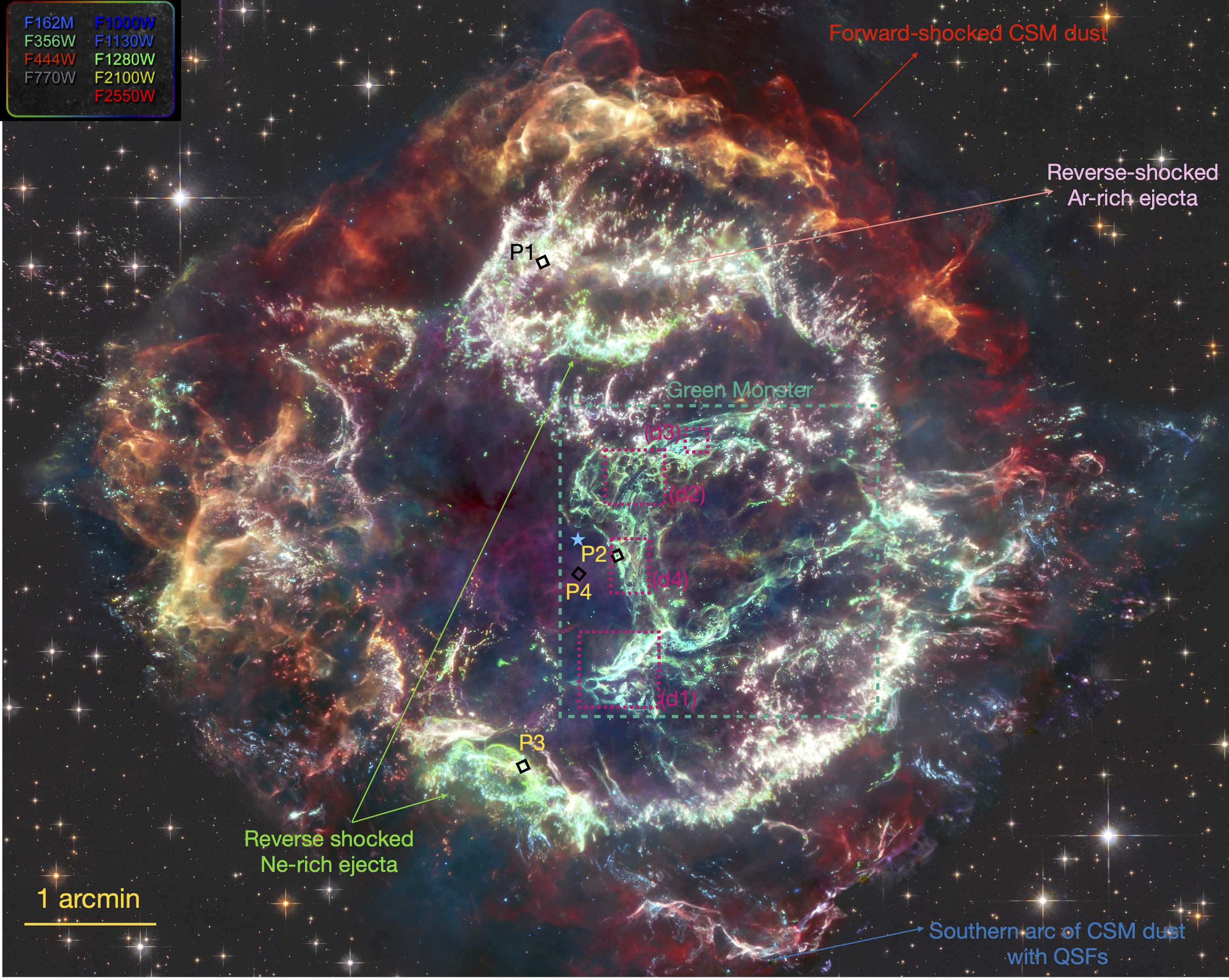}
\caption{Multi-color \textit{JWST}/NIRCam and MIRI image of Cas\,A adapted from \citet{Milisavljevic2024} with the locations of the four MIRI MRS pointings overlaid as black squares. The green rectangular region marks the position of the GM which is shown enlarged in Figure \ref{FigGM}, whereas the magenta rectangles outline several regions of interest of which we provide a multi-wavelength view in Figure\ \ref{FigGMZoom}. The explosion center is highlighted with a blue star. Beyond the coverage of the \textit{JWST} images, we use a 2023 epoch {\sl HST} image to fill in the emission in the background. The legend outlines the representation for each color in this multi-color image. The prominent green emission observed from the GM originates mostly from the MIRI/F1280W filter with contributions from MIRI/F1130W. The GM emission is not to be confused with the reverse-shocked Ne-rich ejecta visible in light green (representative of the bright Ne line in the F1280W filter). The reverse-shocked Ar-rich ejecta are instead mostly colored pink and white thanks to the Ar line contributions to the F770W filter and the dust continuum emission with a characteristic 21\,$\mu$m peak significantly contributing to the F2100W and F2550W images. The forward-shocked CSM dust in the north, east and west is colored red and orange owing to the circumstellar dust emission peaking in the F2100W and F2550W filters. }
\label{FigWebbJudy}%
\end{figure*}

\begin{figure*}
\centering
\includegraphics[width=0.95\textwidth]{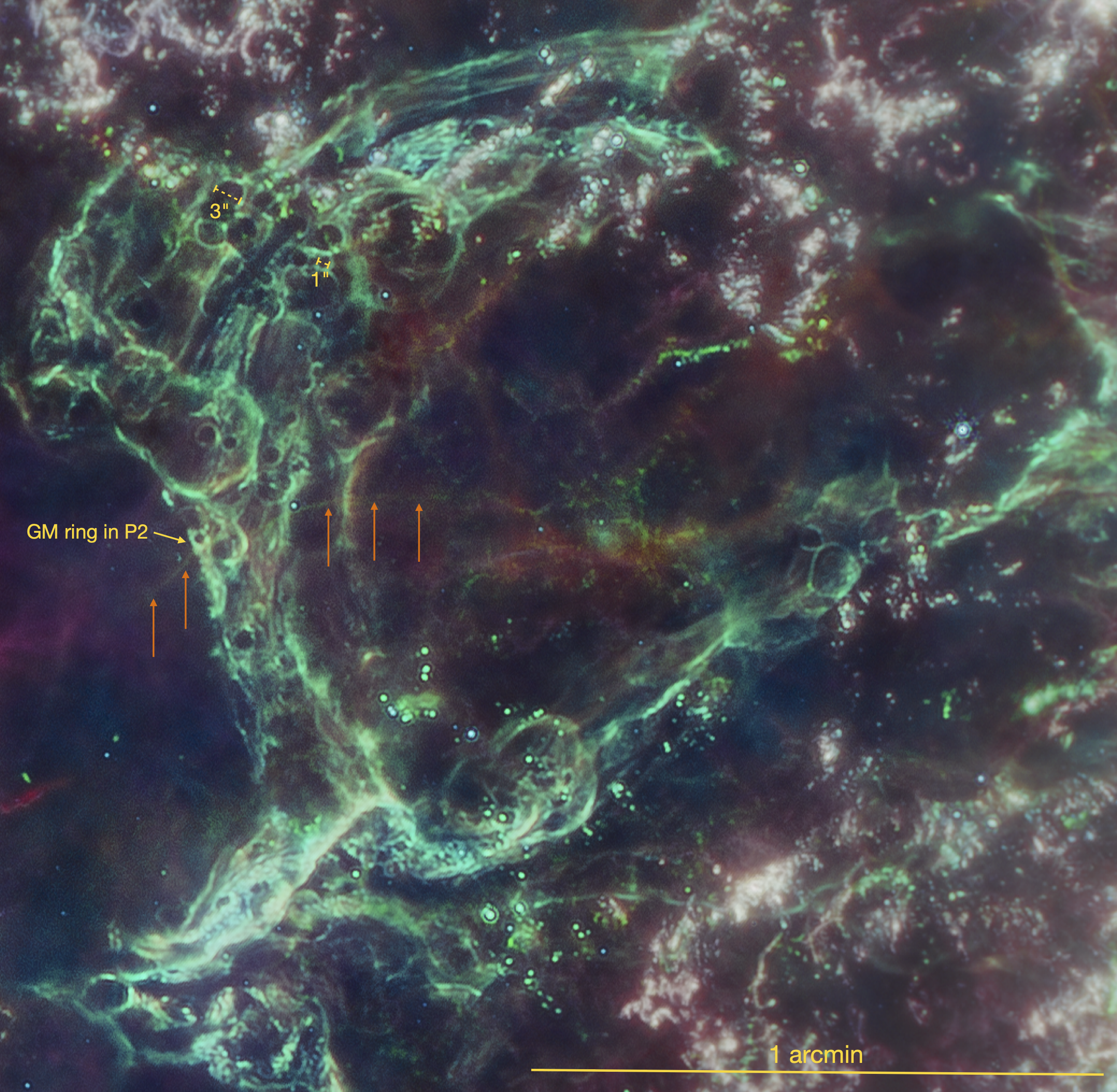}
\caption{This \textit{JWST} multi-color image presents the main GM region applying the same color scheme for NIRCam and MIRI images as used in Figure \ref{FigWebbJudy}. The most striking features are dozens of circular holes speckled across the GM, with sizes ranging between $\approx 1''-3''$ (see the scale bars for some of the holes in the top). The orange arrows point to a very faint and curved $\sim$1 pc long ejecta filament that seems to run up against the base of a ring structure with targeted MRS observations in Position P2 (see also Figure \ref{FigWebbJudy}). We argue that the ejecta interaction with a reverse shock at this specific location is coincidental. }
\label{FigGM}%
\end{figure*}

\subsection{\textit{JWST}/MIRI MRS and \textit{JWST}/NIRSpec spectroscopic data}

Four different regions of interest were zoomed into with \textit{JWST}'s spectroscopic instrumentation (see Figure\,\ref{FigWebbJudy}). One deep drilling was done of a central region in the unshocked ejecta (P4); these observations are discussed in more detail in \citet{Milisavljevic2024}. Three other pointings target reverse-shocked ejecta knots in the north-east with bright Ar line emission (P1), a hole with a surrounding ring in the GM structure (P2), and a reverse-shocked region in the South with bright Ne emission and the presence of bright warm CO gas (P3, see Figure\,\ref{FigWebbJudy}). In this paper, we will present the spectroscopic observations of one circular ring structure in the GM (P2) along with the photometric observations of the entire GM. 

\begin{figure*}
\centering
\includegraphics[width=1.0\textwidth]{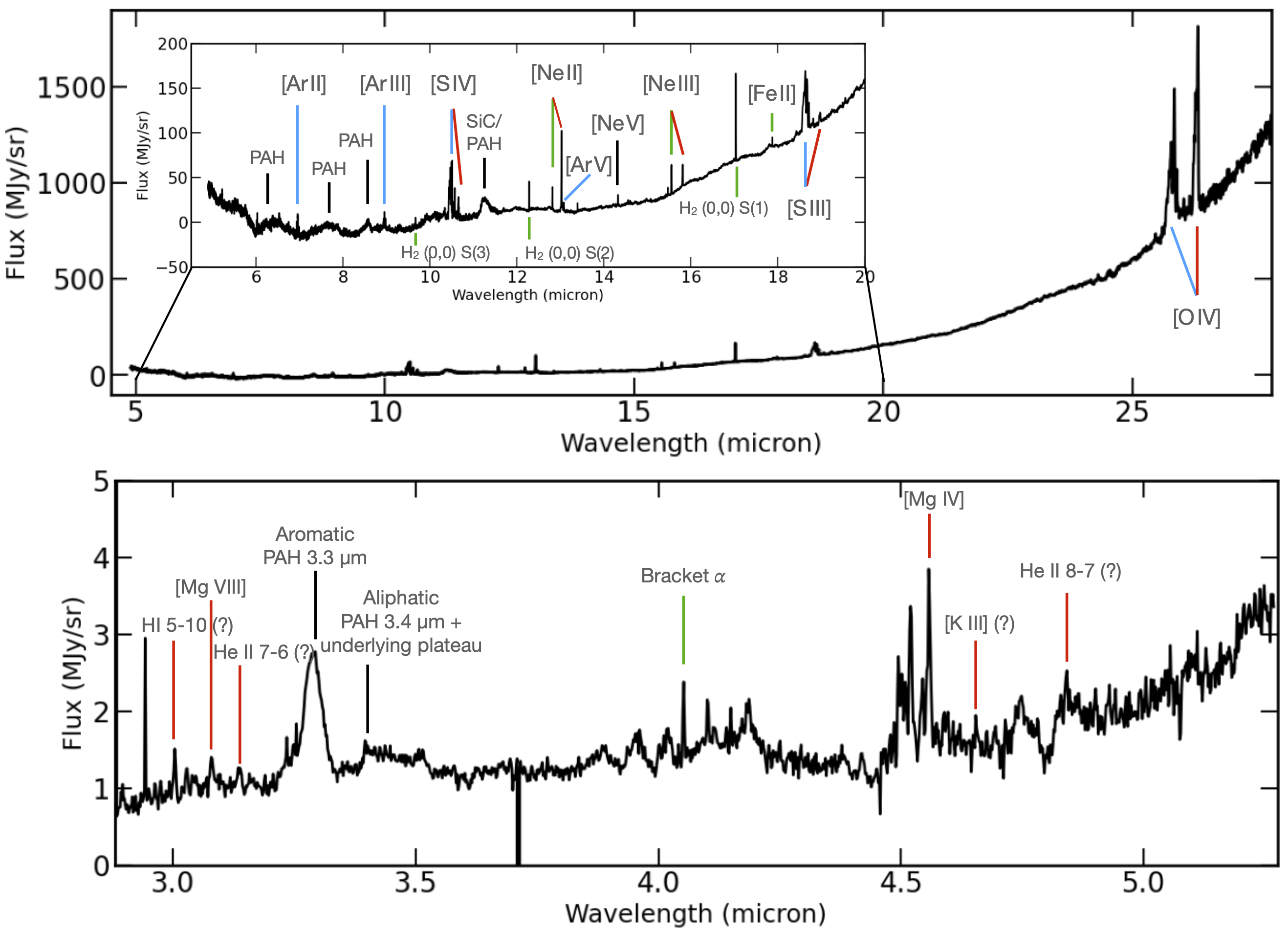}
\caption{Total integrated MIRI MRS and NIRSpec spectra for one specific ring in the GM (Position P2) with line identifications, some of which remain uncertain. The colors indicate whether the line emission is mostly blueshifted (blue), near rest-frame velocity (green) or redshifted (red). Black lines point to specific dust features. For NIRSpec, the flux was integrated in a $2.5\times2.5$ arcsec$^{2}$ box centered on the ring to avoid noisy signal at the edges. In the MRS data, several lines ([Ne\,{\sc{ii}}], [Ne\,{\sc{iii}}]) originate from two distinct velocity (v$=0$ and v$=5000$ km s$^{-1}$) components. The [Ar\,{\sc{ii}}], [Ar\,{\sc{iii}}], [Ar\,{\sc{v}}], [S\,{\sc{iii}}] and [S\,{\sc{iv}}] emission lines mostly originate from blueshifted components, whereas the [O\,{\sc{iv}}] line has multiple blue- and redshifted components. The H$_{2}$ lines trace the ISM along the same line of sight. In the NIRSpec data, the Br\,$\alpha$ line is the only line originating from the v $=0$ km s$^{-1}$ component, whereas the other lines appear to be emitted from a high-velocity (v$=5000$ km s$^{-1}$) component.}
\label{FigSpectra}%
\end{figure*}

Each of these regions was covered with the MIRI Medium Resolution Spectrograph (MRS) and the NIRSpec IFU\footnote{For the long exposure of the unshocked ejecta, the NIRSpec fixed slit mode was used instead of the IFU.} instruments. For specific details on the MIRI MRS data reduction, we refer to the survey paper \citep{Milisavljevic2024}. We briefly summarize the spectroscopic observations for position P2 here, which were taken in November 2022. MRS observations of three different gratings (SHORT, MEDIUM and LONG) guaranteed coverage of the full wavelength range between 4.9 and 27.9\,$\mu$m. The FASTR1 readout mode was used and a 4-point dithering scheme for extended sources was adopted. Each exposure in a given grating setting and wavelength range (MRSSHORT or MRSLONG) consisted of 50 groups and 4 integrations (due to the 4-point-dithering), resulting in an exposure time of 555\,s. Simultaneous imaging was done in the F560W, F770W and F1500W filters. In addition to the standard data reduction pipeline (we used v\,1.11.0), the \texttt{jwst.residual$\_$fringe.ResidualFringeStep} was used and a dedicated background was subtracted using the master background subtraction method. The 12 individual spectra from the different channel and grating combinations were additively scaled to obtain a smooth stitched spectrum. To check the astrometry, we integrated the line-free continuum in various MIRI broadband filters and compared these images to the MIRI mosaics which have been anchored to the GAIA DR3 targets. We found an offset between the MRS data and the MIRI images of $\sim$0.25$\arcsec$ and corrected the WCS of the MRS datacubes for this offset. 
Figure \ref{FigSpectra} (top panel) shows a global extraction of the mid-infrared emission in the dedicated MRS observations of position P2. It is immediately evident from the spectrum that several emission lines --- in particular the [S\,{\sc{iv}}]\,10.51\,$\mu$m, [Ne\,{\sc{ii}}]\,12.81\,$\mu$m, and [O\,{\sc{iv}}]\,25.89\,$\mu$m lines --- have multiple velocity components corresponding to different physical regions along the line-of-sight. 

The NIRSpec IFU data were taken during November and December 2022 with the medium resolution grating G395M (R$\sim$1000) and filter F290LP covering the 2.87$–$5.10\,$\mu$m wavelength range. Two different exposures were taken -- one with and without leakcal -- to allow for corrections for spectral contamination from malfunctioning open MSA shutters in post-processing. Each exposure consisted of 15 groups and 4 integrations due to the 4-point-dithering scheme, resulting in a 933.69\,s exposure time. The NRSIRS2RAPID readout pattern was used. The NIRSpec data were processed with the standard data reduction pipeline (calibration files v\,1.11.4). The spectrum extracted from a $2.5\arcsec\times2.5\arcsec$ box centered on this specific ring in the GM is shown in Figure\,\ref{FigSpectra} (bottom panel). The detected lines are quite faint, making the line identification more difficult. The Br\,$\alpha$ line is near zero velocity, whereas the other lines (i.e., [Mg\,{\sc{viii}}], [Mg\,{\sc{iv}}], He\,{\sc{ii}} 7-6 and He\,{\sc{ii}} 8-7) appear to be emitted from a high-velocity ($v=5000$ km s$^{-1}$) component. Unconfirmed line identifications have been indicated with a question mark. 

\subsection{Ancillary data products}
We use Hubble Space Telescope ({\sl HST}) data from the HST ACS/WFC program ``A Multi-Bandpass ACS Survey of Cassiopeia A: Keeping Up with its Rapidly Evolving Structure" (PI: R. Fesen) targeting Cas\,A in December 2022 and May 2023 using different filters (F450W, F625W, F775W and F850LP) to distinguish between O-Ne-rich and S-Ar-Ca-rich ejecta in Cas\,A. In this paper, we made use of the ACS/WFC F625W image that is sensitive to H$\alpha$ line emission from circumstellar material and to the emission from the 
[\ion{S}{2}]  6716,6731 \AA \ doublet originating from S-rich ejecta. 

\section{Observational characteristics of the Green Monster}
\label{GMobs.sec}
In this section, we review observations that will best indicate the nature of the GM in Section \ref{GM.sec}.

\subsection{Morphology of the Green Monster}
Figure \ref{FigWebbJudy} shows a multi-color MIRI+NIRCam image of Cas\,A highlighting the complexity of this supernova remnant system. The red and orange clouds outside of the main remnant correspond to the material that was lost by the progenitor star before its explosion, and is currently being overrun by the supernova blast wave. The magenta, pink, and white colors reveal clumpy ejecta knots that have been shocked by the remnant's reverse shock. This reverse shocked material also outlines the fast-moving ejecta that form the jet in the East, and the protrusion of the ring in the West. 

Some of the ejecta is particularly rich in neon with prominent emission lines in the MIRI/F1280W filter which makes these regions appear green. Towards the center, the green colors (representative of the emission in the MIRI/F1130W and F1280W filters) reveal a new GM structure that is mostly located within the borders of the supernova remnant. Using \textit{Spitzer} data, \citet{Arendt2014} did identify the brightest part of the GM as the South Spot dust, and speculated it could originate from circum- or interstellar material. An outline of the fainter GM emitting regions is also shown in Figure 4 of that paper, but back then it was not yet identified as belonging to the same structure. Hints of emission from the GM are also seen in earlier \textit{ISO} data \citep{Lagage1996}, suggesting that the GM has been bright at mid-infrared wavelengths for at least $\sim$30 years. 

Zooming into this new structure, we unexpectedly discovered numerous partial or complete circular structures (``holes'') across much of the GM. Figure \ref{FigGM} showcases multi-color MIRI images that provide an enlarged view of these circular structures. From this high-resolution view,  most of
these
features correspond to regions empty, or nearly so, of any infrared emission surrounded by 
thin circular rings of enhanced emission some $0.25'' - 0.5''$ thick.

In an attempt to characterize these circular structures, we first manually identified them in the \textit{JWST}/MIRI F1130W image  
and then measured the range of typical sizes of these structures. From our visual inspection, we identified more than two dozen circular structures with typical sizes ranging between $\sim 1''- 3''$ ($\sim 0.015-0.05~$pc) in diameter. While some ring-like features
have diameters that exceed $3''$, most of these are faint or partial and thus more difficult to 
characterize due to their fainter and/or partial appearance.  
The nearly perfect circular shape of a large fraction of these holes does not leave many possible scenarios to explain their creation, a topic
which we will address in 
Section \ref{GM.sec}.

The dust emission that dominates the GM's emission in the MIRI images almost never traces out a complete ring. Instead, we see that the partial ring structures are characterised by small regions of bright emission, which could indicate that these are local density enhancements in the GM rings, or that the dust is locally heated to higher temperatures boosting its emission. Similar to these bright enhancements in the circular structures, the GM emission is dominated by filamentary and clumpy structures of which some are a lot brighter than the surrounding regions. 

\begin{figure*}
\centering
\includegraphics[width=1.0\textwidth]{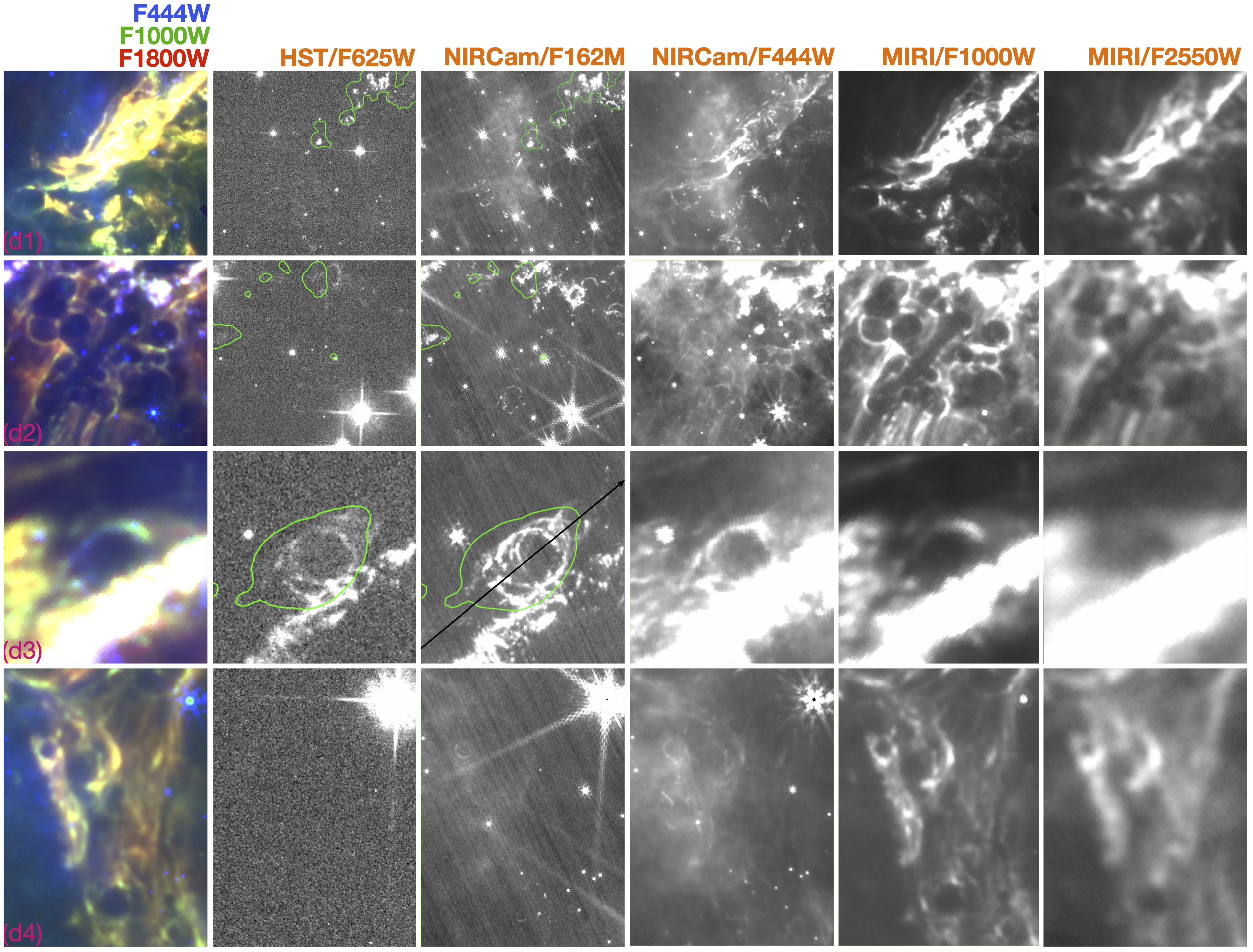}
\caption{Multi-wavelength view on the GM zoomed-in on a few specific regions of interest (see the boxes in Fig\,\ref{FigWebbJudy} for their specific locations). The first column shows RGB images (blue: NIRCam/F444W; green: MIRI/F1000W; red: MIRI/F1800W) of each position, with the subsequent columns presenting the HST/F625W, NIRCam/F162M, NIRCam/F444W, MIRI/F1000W, and MIRI/F2550W images. The HST/F625W and NIRCam/F162M images have the contours of previously identified dense circumstellar clumps (QSFs) in deep [Fe\,{\sc{ii}}]+[Si\,{\sc{i}}] \citep{Koo2018} imaging overlaid in green.}
\label{FigGMZoom}%
\end{figure*}

While the GM structure was first identified from the \textit{JWST}/MIRI images, the \textit{JWST}/NIRCam images also reveal some traces of the GM emission. In the NIRCam images, a diffuse emission component traces non-thermal synchrotron radiation, which is most visible in the NIRCam/F444W image  (see Fig.\ \ref{FigGMZoom}).
The dominant emission in the NIRCam/F162M image originates from bright clumps that trace [Fe\,{\sc{ii}}]\,1.644\,$\mu$m line emission with a possible non-negligible contribution from synchrotron emission (see Appendix \ref{LineFit.sec}). While the NIRCam/F162M filter covers both the [Fe\,{\sc{ii}}]\,1.644\,$\mu$m and 
[Si\,{\sc{i}}]\,1.645\,$\mu$m emission lines, \citet{Koo2018} has argued using previous ground-based narrow-band imaging that the diffuse emission component likely originates from [Si\,{\sc{i}}]\,1.645\,$\mu$m line emission. The [Fe\,{\sc{ii}}]\,1.644\,$\mu$m line emission was found to be more clumped and likely originates from shocked material, which is also consistent with the detection of H$\alpha$ 0.656\,$\mu$m line emission from several Fe-bright CSM clumps (see the second column in Figure\,\ref{FigGMZoom}). 

Some of the Fe-bright clumps also emit in the NIRCam/F444W image, where the emission is likely dominated by non-thermal radiation and/or thermal emission from hot carbonaceous grains. The brightest emission filaments in the NIRCam/F444W image also correspond to the location of the dust emission visible in the MIRI images. However, most of the ring structures are less prominent in the NIRCam images (mostly absent in the F162M image and fainter in the F356W and F444W images), which confirms that the emission in the GM is largely dominated by dust emission across all MIRI filters. The green colors that outline the boundaries of several of these holes in the RGB image (see the first column of Figure \ref{FigGMZoom}) reveal that the dust temperatures in these rings are high compared to the rest of the GM emission and/or that dust properties are different. 

\begin{figure*}
\centering
\includegraphics[width=1.0\textwidth]{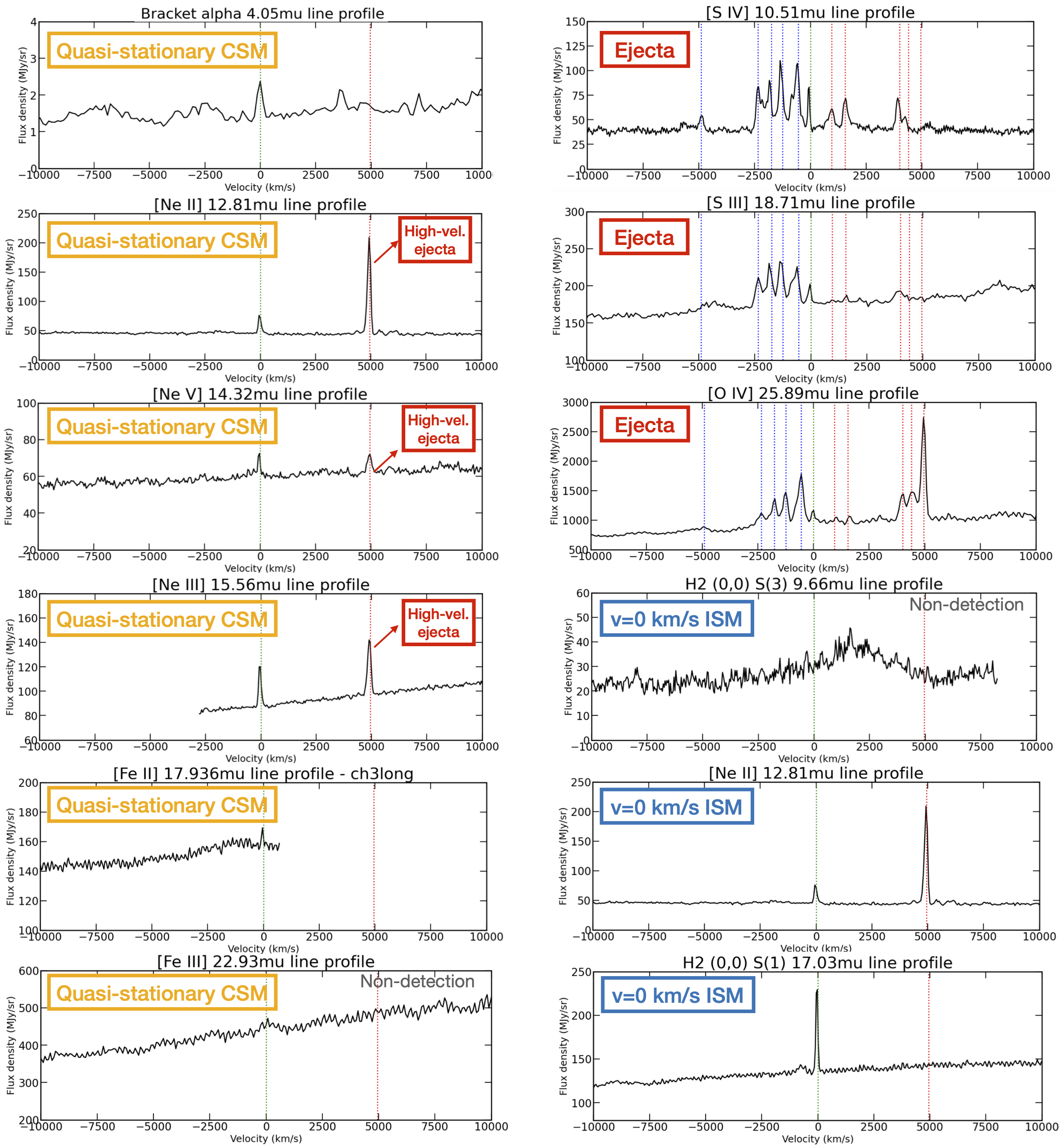}
\caption{Line profiles for the brightest lines observed with NIRSpec and MIRI/MRS for the specific ring in the GM located in Position P2 with the velocity on the x-axis, calculated relative to the rest-frame wavelength in vacuum for each of the transitions (see Table \ref{Line_fluxes}) and the line flux density (in units of MJy/sr) on the y-axis. The spectra were grouped depending on whether the emission originates from quasi-stationary ($v$ near 0 km s$^{-1}$) circumstellar material (yellow box), high-velocity ejecta material (red box) or interstellar material (blue box). In case of the neon lines, there is a contribution from quasi-stationary circumstellar material near the rest-frame velocity and high-velocity ejecta material near 5000 km s$^{-1}$. The vertical dotted lines indicate different velocity components with blueshifts of -4900, -2350, -1750, -1250, -550 km s$^{-1}$ (in blue), a rest-frame velocity of 0 km s$^{-1}$ (in green) and redshifts of +950, +1550, +4000, +4400, +4950 km s$^{-1}$ (in red) in the top three panels in the right column. The other panels have vertical lines overlaid to highlight velocity components with a rest-frame velocity of 0 km s$^{-1}$ (in green) and a redshift of +4950 km s$^{-1}$ (in red).}
\label{FigLineProfiles}%
\end{figure*}

\subsection{Spectral decomposition and analysis}
\label{origin.sec}
In order to shed light on the nature of the GM, and its position along the line-of-sight, we take a look at the kinematic information available in the MIRI MRS and NIRSpec data (see Fig.\ \ref{FigSpectra}) of the GM ring targeted in position P2 (see Figure \ref{FigWebbJudy}). 

We present an overview of the spectral line profiles in Figure \ref{FigLineProfiles} for several of the brightest lines in position P2, and we discuss the fitting procedure to infer line fluxes and maps in Appendices \ref{LineProfile.sec} and \ref{LineFit.sec}. By performing this line fitting in each spaxel, we produced line intensity maps for the most prominent velocity components (see Figures \ref{FigLineMapsRing} and \ref{FigLineMapsFil}). To briefly summarize our results, we find that the [O\,{\sc{iv}}], [S\,{\sc{iv}}] and [S\,{\sc{iii}}] lines have several velocity components tracing different supernova ejecta filaments along the line of sight. In most cases, there is perfect agreement between the velocities of these various line emission components. This suggests that a contribution from the [Fe\,{\sc{ii}}]\,25.99\,$\mu$m line --- which would show up as an extra line emission component shifted 0.1\,$\mu$m with respect to the [O\,{\sc{iv}}] line ---  is minimal. The absence of any other Fe lines makes a significant contribution from the [Fe\,{\sc{ii}}]\,25.99\,$\mu$m line unlikely. There is an interesting $v_{\text{rad}}\sim5000$ km s$^{-1}$ filament that appears bright in [O\,{\sc{iv}}], [Ne\,{\sc{ii}}] and [Ne\,{\sc{iii}}], but does not show any corresponding [S\,{\sc{iv}}] or [S\,{\sc{iii}}] emission (see Fig.\ \ref{FigLineProfiles}). This $v_{\text{rad}}\sim5000$~km~s$^{-1}$ O-rich filament touches the base of the ring structure in position P2 (see Fig.\ \ref{FigLineMapsFil}), and we will discuss the origin of its emission in Scenario II of Section \ref{GM.sec}. Aside from this high-velocity component, there is [Ne\,{\sc{ii}}] and [Ne\,{\sc{iii}}] emission corresponding to a $v_{\text{rad}}\sim0$ km s$^{-1}$ velocity component originating from the eastern part of the GM ring in this location (see Figs.\ \ref{FigLineProfiles} and \ref{FigLineMapsRing}). Apart from the Ne lines, faint emission near $v_{\text{rad}}\sim0$ km s$^{-1}$ is also detected for the [Fe\,{\sc{ii}}]\,17.94\,$\mu$m and hydrogen Br\,$\alpha$ $n=5\rightarrow n=4$ recombination lines. Table \ref{Line_fluxes} presents an overview of the line fluxes that were inferred for this near rest-frame velocity component. 

Through convolution of the extracted spectra with and without line emission with the filter response curves, we are able to estimate the contribution of various lines to the MIRI broadband filters. We find that the average contribution of lines to the MIRI filters is limited, amounting to a few $\%$ at most with the exception of the [O\,{\sc{iv}}] 25.89\,$\mu$m line which significantly contributes to the \textit{JWST}/MIRI F2550W filter ($\sim$$16\%$). Assuming that the MIRI spectral data for this specific position in the GM is representative, we can conclude that the GM emission is dominated by dust continuum emission rather than line emission. 

The complex kinematic structures along the line-of-sight reveal two velocity components (see Fig.\ \ref{FigLineProfiles}) with line emission resembling part of the dust continuum emission of the GM probed in the MIRI broadband filters.  
The dust emission in the ring structure on the east side is reflected in the [Ne\,{\sc{ii}}], [Ne\,{\sc{iii}}], [Fe\,{\sc{ii}}]\,17.936\,$\mu$m and Br\,$\alpha$ line emission centered around velocities of $-50$ to 0 km~s$^{-1}$, and also corresponds to the morphology of the [Fe\,{\sc{ii}}]\,1.644\,$\mu$m line emission in the NIRCam/F162M images (see Figure \ref{FigZoomRing}). On the other hand, the southern bright part of the ring is co-spatial with the position of a bright O-rich ejecta filament with bright [O\,{\sc{iv}}], [Ne\,{\sc{ii}}] and [Ne\,{\sc{iii}}] line emission with velocities around 5000 km s$^{-1}$ (see Fig.\  \ref{FigLineMapsFil}). 

In the remainder of this section, we will discuss further observational evidence linked to the near-rest-frame velocity component that shows a spatial correlation with the eastern side of the one specific GM ring targeted with MRS observations that will help us to constrain the nature of the GM in Section \ref{GM.sec}. \\

\subsubsection{Line broadening}
\label{Broadening.sec}
While the line flux maps of the [Ne\,{\sc{ii}}], [Ne\,{\sc{iii}}], [Fe\,{\sc{ii}}]\,17.94\,$\mu$m and Br\,$\alpha$ lines nicely trace the eastern part of the GM ring in Position P2 (first column of Fig.\  \ref{FigLineMapsRing}) detected with MIRI in dust continuum emission, the radial velocities and velocity dispersions of these lines span quite a range of values with substantial positional variations (see Fig.\  \ref{FigLineMapsRing}, second and third columns). The broadening of the [Fe\,{\sc{ii}}]\,17.94\,$\mu$m and Br\,$\alpha$ line profiles appears to be solely driven by the instrumental resolution (i.e., $\sim$100 and $\sim$300 km s$^{-1}$, respectively). For both Ne lines, the brightest blob of emission towards the north-east shows the largest radial velocities and line widths, whereas the fainter region towards the south-east does not show any significant broadening of the lines beyond what is expected from instrumental effects.

We now focus on the brightest blob of emission detected in the north-eastern part of the ring in both Ne lines to study the extent of the line broadening. Correcting for the instrumental resolution --- assumed to be $\sim90-100$ km s$^{-1}$ and $\sim120-130$ km s$^{-1}$ for the [Ne\,{\sc{ii}}] and [Ne\,{\sc{iii}}] lines, respectively, we infer intrinsic FWHM line widths of $v_{\text{FWHM}}\sim$150 km s$^{-1}$ and $v_{\text{FWHM}}\sim$100 km s$^{-1}$ for [Ne\,{\sc{ii}}] and [Ne\,{\sc{iii}}], respectively. The significant broadening of these lines suggests the presence of shocked gas.
However, while the north-eastern part of the ring could be consistent with shocked gas, we cannot tell whether the gas in the remainder of this GM ring is shocked due to the limited spectral resolution. Later on, we will compare the observed line flux ratios with shock models to study whether the emission has been shocked ionised and excited, or whether pre-ionisation of the unshocked material upstream can also contribute to the line emission (see Section \ref{PhotovsShock.sec})

We can now use these line widths to study the gas conditions. Assuming that non-thermal broadening is subdominant, we can estimate an upper limit on the gas temperature using
\begin{equation}
FWHM~[\AA]~=~\lambda_{0}~[\AA]~\left(\frac{8~k_{\text{B}}~T[K]~\ln(2)}{m~c^{2}}\right)^{1/2},
\end{equation}
with the central wavelength $\lambda_{0}$, the Boltzmann constant $k_{\text{B}}$, the speed of light $c$ and the mass $m$ of the particles, in this case equal to 20$\times$ $m_{p}$ for these Ne emission lines with $m_{p}$ the proton mass. We infer the velocity width from FWHM [km/s] = c $\times$ FWHM [$\AA$]/$\lambda_{0}$ [$\AA$], which suggests upper limits on the gas temperatures ranging from $T_{\text{gas}}$ $\sim$ 5$\times10^{6}$ to 10$^{7}$ K. These gas temperatures would suggest shocked gas or post-shock that has been radiatively cooled.

\subsubsection{Photo- or shock ionised material?}
\label{PhotovsShock.sec}
\begin{figure*}
\centering
\includegraphics[width=1.0\textwidth]{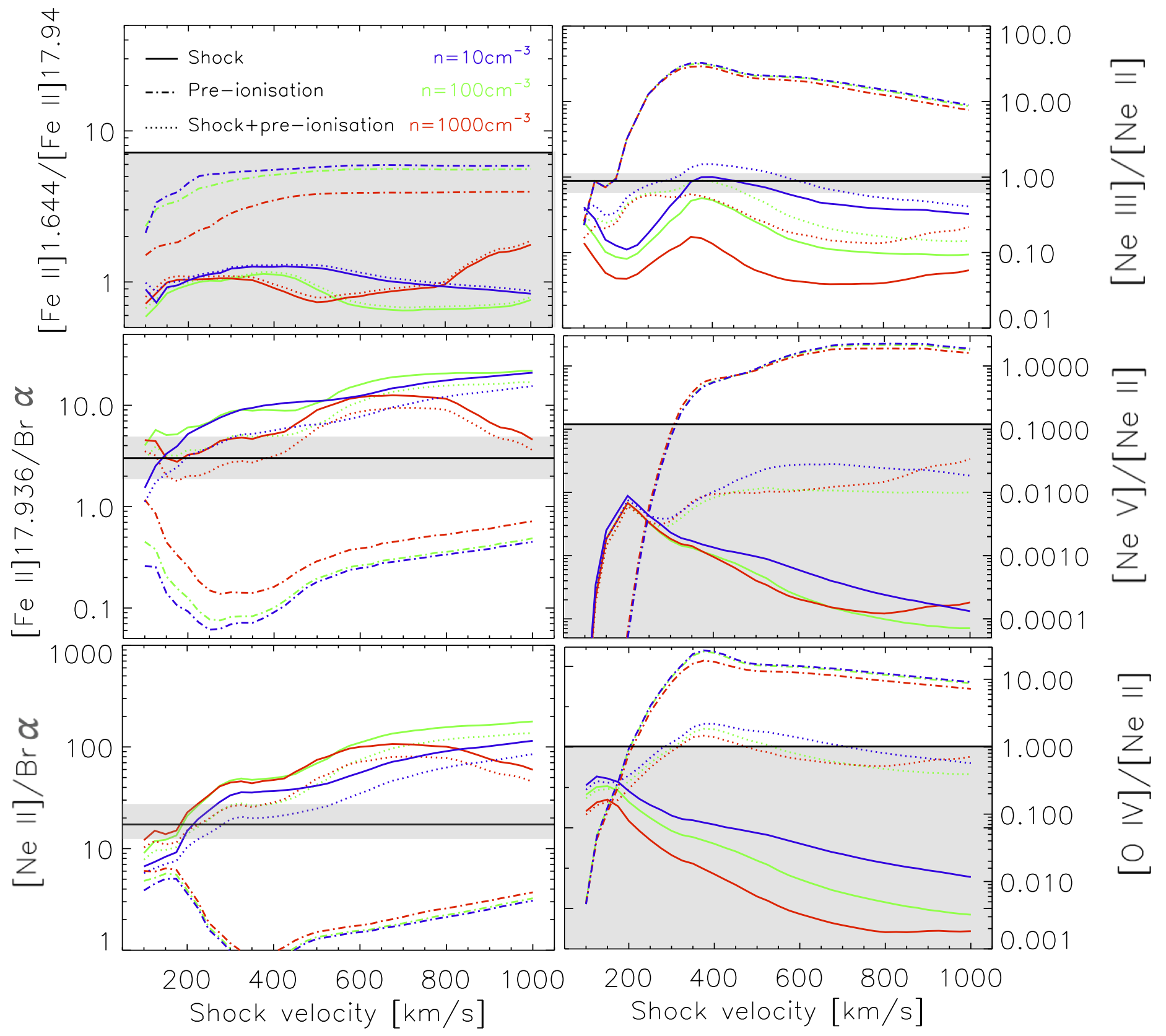}
\caption{Comparison of the observed line ratios (grey shaded regions) with the library of radiative shock models from \citet{Allen2008} for models with different pre-shock densities $n$ of 10 (blue), 100 (green) and 1000 (red) cm$^{-3}$, and different ionisation mechanisms: shock-ionised emission (solid lines), pre-ionisation of unshocked material by the shock's radiation (dashed-dotted lines) or both (dotted lines). The observations are consistent with low shock velocities ($<200$ km s$^{-1}$), but are unable to exactly pinpoint the ionisation and excitation mechanism and the gas density.}
\label{FigMAPPINGs}%
\end{figure*}

With several detected emission lines in the NIRSpec and MIRI MRS pointing coinciding with the specific GM ring structure in Position P2, we attempt to constrain the physical conditions of the GM at that specific location. We focus on the near rest-frame velocity component, since the line emission of a high-velocity component appears unrelated to the GM (see Section \ref{GM.sec}). We show several observed emission line ratios (see horizontal black lines) in Figure \ref{FigMAPPINGs}, and compare them with the estimates from the \texttt{MAPPINGS} III library of fully-radiative shock models \citep{Allen2008} at solar metallicity. The line ratios are depicted as a function of the modeled shock velocity. We overlay the model line ratios for \texttt{MAPPINGS} models with different pre-shock densities $n$ of 10 (blue), 100 (green) and 1000 (red) cm$^{-3}$, and we consider models of shock-ionised emission (solid lines), pre-ionisation of unshocked material by the shock's radiation (dashed-dotted lines) and models that account for both effects (dotted lines).

Comparing the observed line ratios within the error bars (grey shaded regions) with the \texttt{MAPPINGS} models, the observations tend to be reproduced best by models with low shock velocities ($100-300$ km s$^{-1}$), which is consistent with the observed radial velocities of -50 to 0 km s$^{-1}$. For dedicated modelling efforts in future work, we will explore a finer grid of shock velocities, also extending below 100 km s$^{-1}$. The upper limits on [O\,{\sc{iv}}]/[Ne\,{\sc{ii}}] and [Ne\,{\sc{v}}]/[Ne\,{\sc{ii}}] line ratios and the observed [Ne\,{\sc{iii}}]/[Ne\,{\sc{ii}}] line ratio rule out pure photo-ionisation models for high shock velocities ($v >200$ km s$^{-1}$), whereas the observed [Ne\,{\sc{ii}}]/Br\,$\alpha$ and [Fe\,{\sc{ii}}]\,17.94\,$\mu$m/Br\,$\alpha$ line ratios suggest that shock ionisation and excitation by high-velocity shocks ($v >300$ km s$^{-1}$) is unlikely. At these low shock velocities, it is difficult to distinguish between pure shock or pure photo-ionisation models, or a combination of the two. While the observed [Ne\,{\sc{ii}}]/Br\,$\alpha$ and [Fe\,{\sc{ii}}]\,17.94\,$\mu$m/Br\,$\alpha$ line ratios would tend to favor pure shock ionisation models, other line ratios (e.g., [Ne\,{\sc{iii}}]/[Ne\,{\sc{ii}}] and [Fe\,{\sc{ii}}]\,1.644\,$\mu$m/[Fe\,{\sc{ii}}]\,17.94\,$\mu$m) are less conclusive.

We have already seen that some regions show significant line broadening that could be indicative of shocks, while others may be consistent with pure photo-ionising radiation from shocks (see Section \ref{Broadening.sec}). Given that we are likely sampling a variety of conditions within this single GM ring, in addition to the potential effect of uncertain dust extinction corrections for some of the lines (e.g., [Fe\,{\sc{ii}}]\,1.644\,$\mu$m and Br\,$\alpha$, see Appendix \ref{LineFit.sec}), it is not surprising that models and observations are not in perfect agreement. For instance, the [Ne\,{\sc{ii}}] and [Ne\,{\sc{iii}}] line emission arises from two bright blobs within the ring with slightly different line emission ratios. The blob in the north-east has a below average line ratio ([Ne\,{\sc{iii}}]/[Ne\,{\sc{ii}}]$\sim$0.5) consistent with shock ionisation with a potential contribution from pre-ionisation effects, which is consistent with the line broadening suggestive of shock excitation. The south-eastern blob, on the other hand, is characterised by slightly higher line ratios which could imply that photo-ionisation plays a more prominent role.

\subsubsection{Offsets between line and dust emission}

\begin{figure*}
\centering
\includegraphics[width=1.0\textwidth]{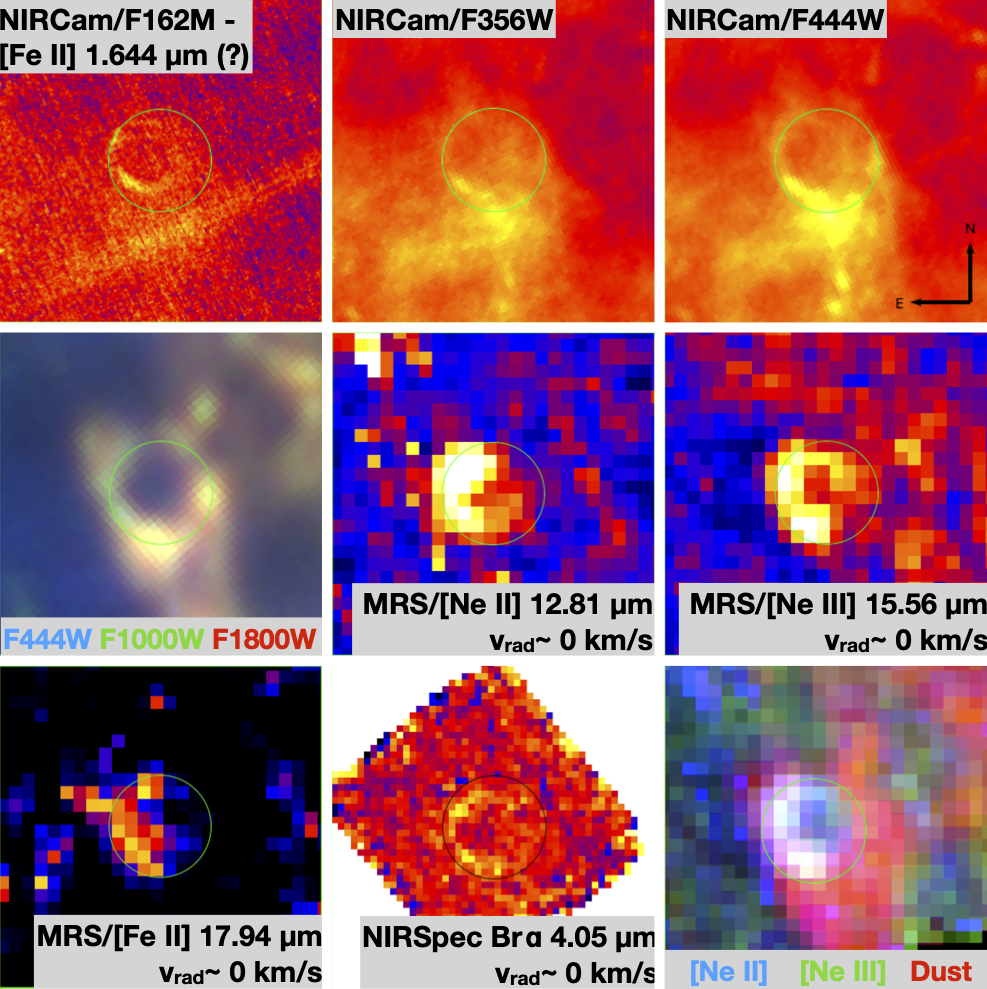}
\caption{Multi-wavelength view on the GM ring in position P2 (see Figure \ref{FigWebbJudy}) showing the NIRCam F162M (PSF FWHM: $0.055\arcsec$), F356W (PSF FWHM: $0.116\arcsec$) and F444W (PSF FWHM: $0.145\arcsec$) images (top row), NIRCam+MIRI RGB image (F444W, F1000W, F1800W with PSF FWHMs of $0.145\arcsec$, $0.328\arcsec$ and $0.591\arcsec$, respectively) and MRS MRS line flux maps for [Ne\,{\sc{ii}}] 12.81\,$\mu$m (PSF FWHM: $\sim0.529\arcsec$) and [Ne\,{\sc{iii}}] 15.56\,$\mu$m (PSF FWHM: $\sim0.619\arcsec$) lines (middle row) and MRS [Fe\,{\sc{ii}}] 17.94\,$\mu$m (PSF FWHM: $\sim0.698\arcsec$) and NIRSpec Br\,$\alpha$ 4.05\,$\mu$m (PSF FWHM: $\sim0.15\arcsec$) line flux maps for the near $v_{\text{rad}}=0$ km s$^{-1}$ velocity component and a multi-color RGB image comparing the dust with the line emission (bottom row). The images are at their native resolution; the FWHM of the PSF has been indicated between parentheses here for guidance. Each of the cut-outs has the same 5 arcsec $\times$ 5 arcsec field of view. The green or black circle with radius of 0.8\,arcsec overlaid on the continuum and line emission maps serves to guide the eye when comparing dust continuum, synchrotron and line emission in this region.}
\label{FigZoomRing}%
\end{figure*}

The rings are visible in dust continuum emission, and spectroscopic observations of a particular ring structure in Position P2 (see Figures \ref{FigZoomRing} and \ref{FigLineMapsRing}) also show that part of the ring structure emits in the [Ne\,{\sc{ii}}] 12.81\,$\mu$m, [Ne\,{\sc{iii}}] 15.56\,$\mu$m, Br\,$\alpha$ 4.05\,$\mu$m, [Fe\,{\sc{ii}}] 17.94\,$\mu$m lines and potentially the [Fe\,{\sc{ii}}] 1.644\,$\mu$m  line in the NIRCam/F162M image. However, the line emission is consistently offset by $0.2-0.4\,\arcsec$ from the dust emission towards the inner parts of the ring structure (see Figure\,\ref{FigZoomRing}). 
While belonging to the same structure, we argue that the offset between line and dust emission could be due to different densities, temperatures and/or ionisaton conditions in the inner and outer parts of the rings. Alternatively, the offsets between line and dust emission could also result from the destruction of dust grains in the wake of the shock, resulting in the detection of line emission from Fe and other elements mostly in the inner ring regions. This last scenario seems to fit nicely with the warm IR colors of the rings in the GM (see the first column of Figure \ref{FigGMZoom}).

\subsection{Blue- or redshifted material?}
Both the resolved line modelling (see Figure \ref{FigLineMapsRing}) and the global extractions (see Table \ref{Line_fluxes}) suggest line velocities between $-50$ and 0 km s$^{-1}$ for the Ne, Fe and H line emission originating from the GM ring in Position P2. But the large velocity dispersion ($100-200$ km s$^{-1}$, see Section \ref{Broadening.sec}) and the moderate velocity resolution ($100-300$ km s$^{-1}$) prevent us from drawing any firm conclusions on the position of this ring and the GM along the line of sight. Since the line emission detected in the MIRI MRS cubes for this specific ring location (P2) suggests that the GM is shocked, the measured radial velocities and internal velocity dispersion may have been affected by shock processing.

However, the significant blueshift (-2300 km s$^{-1}$) of the X-ray emission associated with several GM locations \citep[][]{Vink2024} firmly places the entire GM structure in front of Cas\,A. The difference in radial velocities measured at X-ray and mid-infrared wavelengths suggests that we are tracing distinct parts of the GM material at those wavelengths. The X-ray emission is tracing low density material compared to IR.
The X-ray emission also originates mostly from a hot ($>10^{7}$\,K) plasma \citep{Vink2024}, while the mid-infrared line emission could trace cooler (pre- or post-shock) material in the GM. However, the significant line broadening in some parts of the GM ring (see Section \ref{Broadening.sec}) seems to suggest that temperature differences are rather small.

\section{The nature of the Green Monster}
\label{GM.sec}
In this section, we outline four different scenarios for the origin of the GM, and discuss the probability of each scenario in light of the photometric and kinematic information available from the \textit{JWST} observations and presented in Section \ref{GMobs.sec}. In brief, we inferred that low-velocity ($v_{\text{rad}}=-50$ to $0$ km s$^{-1}$) Ne, Fe and H line emission is co-spatial with part of the GM ring in Position P2, and reveals signatures (i.e., line broadening and shock ionisation) of shocked gas. There is also a high-velocity ($v_{\text{rad}}=+5000$ km s$^{-1}$) filament that appears connected to the base of this GM ring structure. 

In view of these findings, we outline four possible scenarios for creating the GM's circular structures visible in the NIRCam and MIRI images. Specifically, we considered Scenario I: these circular structures are the result of the expansion of 
small Ni-rich ejecta, Scenario II: they result from interactions of high-velocity ejecta filaments with dense circumstellar clumps on the rear side of Cas\,A, Scenario IIIa: they are created through interactions with small, high-velocity ejecta 
''bullets'' \textit{before} the forward shock impact, or  
Scenario IIIb: they form from interactions with ejecta fingers generated by hydrodynamic instabilities introduced at the contact discontinuity between the shocked layers of ejecta and circumstellar material \textit{after} the forward shock impact. 

\subsection{\textbf{Scenario I}:  Ni bubbles}

Mapping out the unshocked ejecta, \citet{Milisavljevic2015} have shown that the remnant's interior has a bubble-like morphology which could be shaped through turbulent mixing processes that encouraged the development of outwardly expanding plumes of radioactive $^{56}$Ni-rich ejecta material. The holes in the GM are reminiscent --- although on much smaller scales --- of these bubbles of radioactive ejecta that were inferred to have been $^{56}$Ni-rich and that have shaped the multiple ring structures and bubble-like morphology in Cas\,A's ejecta \citep{Milisavljevic2015}. Hence, the GM could correspond to ejecta material in this case, and we would expect a substantial amount of the decay product, $^{56}$Fe, in the interior of these cavities. The \textit{JWST} data do not reveal any sign of Fe emission in the interior of the GM rings. Instead, Fe emission is only found in the GM rings in the form of [Fe\,{\sc{ii}}] 1.644\,$\mu$m line emission traced in the NIRCam/F162M image and the [Fe\,{\sc{ii}}] 17.936\,$\mu$m line emission detected in GM ring covered with MIRI MRS. 

In this $^{56}$Ni bubble scenario, the GM would be located in the central regions of the remnant dominated by unshocked ejecta material. We have seen that the line broadening and shock excitation conditions of the mid-infrared lines imply that some of the GM material must originate from shocked gas, making this $^{56}$Ni bubble scenario to be very low probability. The small size of the rings are also difficult to explain in this scenario, which would require a significantly slower expansion rate or recent expansion events in comparison to the bigger $^{56}$Ni bubbles that have structured the inner ejecta of Cas\,A on much larger scales \citep[][]{Milisavljevic2015}. 
Furthermore, the prominent mid-infrared dust emission is unlikely to originate from the unshocked ejecta since there are no radiative heating mechanisms that could heat the dust to such high temperatures ($T_{\text{dust}}=130-300$\,K, De Looze et al.\,in prep.). 

\subsection{\textbf{Scenario II}: Interaction between high-velocity ejecta filaments and dense circumstellar knots}
In this scenario, we discuss two possible interpretations under the assumption that the observed connection between the high-velocity ejecta filament and the GM ring in position P2 is not a mere coincidence due to projection effects but rather results from a physical link implying an interaction between a high-velocity ejecta filament and the GM material.

First, we postulate that the GM could result from interactions between high-velocity ejecta filaments and dense CSM knots. Instead of holes, the ring structures would correspond to dense CSM clumps that are currently impacted by Cas~A's high-velocity ($\sim$5000 km s$^{-1}$) O, S, Ar-rich ejecta filaments (see Figure \ref{FigLineMapsFil}). 
Upon impact the high-velocity material continues its path most easily around the dense CSM knots. Zooming out to have a wide-angle view, we can see that there is indeed a large-scale, $>$1 pc in size, ejecta filament that touches upon the GM ring targeted in MIRI MRS observations (see Figure \ref{FigGM}). The increased intensity of the [O\,{\sc{iv}}]\,25.89\,$\mu$m line around 5000 km s$^{-1}$ suggests that an interaction is taking place at that specific position corresponding with the bottom part of this ring structure (see Figure \ref{FigLineMapsFil}). If we assume that this high-velocity ejecta filament is hitting a high density circumstellar clump, these clumps would need to be optically thick beyond 30\,$\mu$m and/or the dust would need to be very cold to prevent emission in the \textit{JWST} MIRI filters. However, the absence of any large-scale filaments coinciding with other rings in the GM make this an improbable scenario. It is furthermore unclear how the interaction of a high-velocity ejecta filament could create ring structures as observed in the GM -- it would be more likely for an arc-like structure to form in this case.

Assuming ballistic motion of this high-velocity ($v=5000$ km~s$^{-1}$) oxygen-rich ejecta filament, we estimate that an impact occurs at a distance of about $\gtrsim$1.79\,pc from the explosion center for an explosion that occurred 350 years ago. This position corresponds with the average position of the reverse shock estimated in the plane of the sky \citep{Vink2022}. Assuming the evolution of the forward and reverse shocks has progressed in a similar way on the rear side of the remnant, this high-velocity ejecta filament could be traversing the reverse shock. In this case, the GM could correspond to a sheet of high-velocity ejecta material that is currently encountering the reverse shock. The holes would be created by dense CSM knots that must have survived the blast wave, and now happen to lie in the vicinity of the reverse shock. 

The interaction between the ejecta material and the dense CSM knots would produce low-velocity line emission on the surface of the CSM clumps, in which case the observed rings would correspond to the limb-brightened surfaces of the CSM clumps. Since we only observe low-velocity ($v_{\text{rad}}=$-50 to 0 km s$^{-1}$) line emission co-spatial with the GM ring targeted in MRS observations and no velocity information linked to the whole of the GM structure is available, we cannot rule out that the GM dust-emitting material is moving at high velocities ($\sim$5000 km s$^{-1}$). However, we consider it unlikely that the GM would correspond to high-velocity ejecta material that is currently hit by the reverse shock. Shock ionisation upon traversing the reverse shock makes ejecta knots emit brightly in mid-infrared lines \citep{Docenko2010,DeLaney2010} along with bright dust emission \citep{Rho2008,Arendt2014}, whereas the GM appears to be mostly dominated by dust continuum emission. The absence of any O, S or Ar lines in the GM ring within position P2 suggests that line emission is negligible, at least in this region. We therefore consider this scenario to be unlikely.   

Instead, we argue that the interaction of the high-velocity ejecta filament with the reverse shock is unrelated to the GM and happens coincidentally on the same line-of-sight. 
If we compare the emission of this specific GM ring to similar ring structures observed with broadband \textit{JWST}/MIRI filters in the GM, it also becomes evident that the emission in the (partial) ring surrounding the empty hole is much more pronounced (see Figure \ref{FigZoomRing}), in particular at the bottom, which suggest that the nature of this particular ring may be different. Thus, we argue that part of the ring structure towards the base of the ring in this MIRI MRS pointing is contaminated by high-velocity ejecta impacting the reverse shock on the rear side of the remnant. 

\subsection{\textbf{Scenario III:} Forward-shocked circumstellar dust with holes created by ejecta interactions}
In our preferred scenario, we argue that the dust-emitting GM corresponds to circumstellar material that has been impacted by the forward shock, and the circular structures are created by interactions with ejecta material. 
Several lines of evidence support this scenario. The small-scale structure of the GM resembles the circumstellar dust seen towards the north and south-west of Cas\,A. In comparison, shocked ejecta filaments are composed of small knots and ejecta shrapnel that formed in the aftermath of the explosion, which are not visible in the GM. In addition, the similar shape of the mid-infrared dust emission spectrum of the GM and the circumstellar material seen in projection towards the north-east and south of Cas\,A (De Looze et al.\,in prep.) suggests that the dust composition is similar, hinting at the same chemical make-up of the material and similar pathways that have led to dust formation.

There is also a correlation between the X-ray emission and the dust emission in the GM \citep{Vink2024}, which supports the shock interaction scenario. An X-ray analysis of the GM \citep{Vink2024} further corroborates this conclusion. X-ray spectral extractions across different regions of the GM show consistently blue-shifted emission corresponding to shock velocities of -3500 km s$^{-1}$. These shock velocities are representative of the forward shock speeds measured for Cas\,A in the plane of the sky (5000-7000 km s$^{-1}$, \citealt{Vink2022}) --- although somewhat on the low side, and reveal that the CSM in the GM is likely interacting with the forward shock on the near side of the supernova remnant. The presence of dense circumstellar material in front of Cas\,A would have caused the forward shock to slow down, which is consistent with the measured slow forward shock speeds. 

\begin{figure*}
\centering
\includegraphics[width=0.99\textwidth]{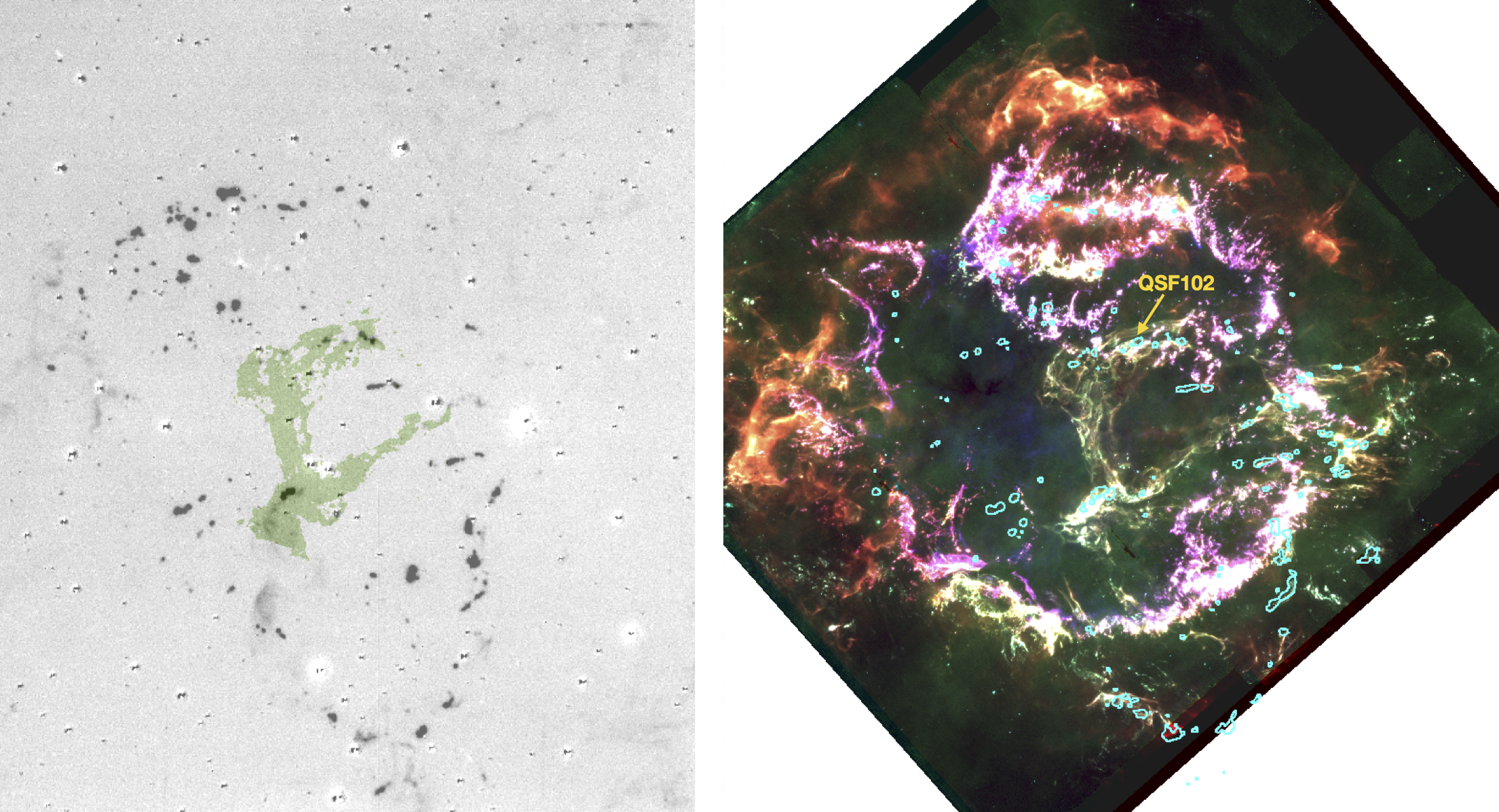}
\caption{\textit{Left:} The locations of dense CSM clumps, often called quasi-stationary flocculi (QSFs), as traced by H$\alpha$ emission. The figure was adapted from \citet{Fesen2001} to show a green outline of the location of the GM, which nicely fits in the middle of the formerly identified dense CSM knots. \textit{Right:} Multi-color MIRI image (blue: F1000W, green: F1130W, red: F2550W) with contours indicating the locations of these dense QSFs as identified by \citet{Koo2018}. }
\label{FigQSFs}%
\end{figure*}

Furthermore, we find a match between the position of the GM, and the location of a handful of dense circumstellar clumps, the so-called quasi-stationary flocculi (QSFs). In Figure \ref{FigQSFs}, we overlaid the positions of the more than 50 QSFs identified from optical and near-infrared data \citep{Koo2018} on a multi-color \textit{JWST}/MIRI image (right panel). A high-resolution comparison is shown in Figure \ref{FigGMZoom} (see the contours in the second and third column) for some specific regions. These figures reveal that some of the QSF positions align well with the GM's dust emission. The broadening (100-200 km s$^{-1}$) of the observed line emission in the GM ring targeted with MIRI MRS is furthermore consistent with the significant broadening ($\sim$200 km s$^{-1}$) observed for shocked QSFs around Cas\,A \citep{Koo2020}, which is in sharp contrast with the expected line width of unshocked CSM material ($\sim$ 8 km s$^{-1}$), and suggests that some GM rings host dense shocked CSM material.

However, it is important to note that there is not a perfect one-to-one correlation between the QSFs and the rings detected in the GM. 
We will return to a discussion on the origin and connection to the various CSM components in more detail in Section \ref{CSM.sec}. 

Finally, an important clue comes from the line emission that is associated with the dust-emitting ring structure. We observe the emission of Ne- and Fe-rich material at near rest-frame velocities ($-50$ km s$^{-1}$ to 0 km s$^{-1}$) tracing the eastern side of the ring structure (see Figure \ref{FigLineMapsRing}). In the NIRSpec data, we also observe hydrogen Br\,$\alpha$ line emission near 4.05\,$\mu$m, which suggests that the material is enriched with hydrogen. This scenario is further corroborated by the detection of H\,$\alpha$ emission coinciding with several dense Fe-rich clumps in the GM (see Figure\,\ref{FigGMZoom}). Given that Cas\,A resulted from a Type IIb explosion, we expect little hydrogen to be present in the ejecta. Hence, the small radial velocities and the detection of hydrogen, neon and iron emission lines perfectly fit the picture of circumstellar material ejected 3$\times$10$^{4}$ yrs to 10$^{5}$ yrs before the SN explosion (calculated for wind velocities ranging from 30 to 100 km s$^{-1}$). 

While we are confident that the GM corresponds to circumstellar material, the exact details on the mechanism capable of creating the circular rings and holes are still missing. We outline two possible scenarios (IIIa and IIIb) which both involve CSM interactions with especially high-velocity small ejecta knots to create the observed circular rings. 

\subsubsection{\textbf{Scenario IIIa}: Fast-moving ejecta knots pierce holes \textit{before} forward shock impact}
Here we postulate that small, fast-moving ejecta bullets have pierced holes in a thin sheet of CSM material, which would create a shock wave moving radially through the CSM layer such that it creates a hole that becomes larger over time (in the absence of any external pressure source such as a new shock impact). 
In this view, the creation of these holes would require ejecta knots that have been accelerated to very high velocities in the aftermaths of the explosion and, in fact, such high-velocity ejecta knots have been identified around Cas~A \citep{Fesen2001, Fesen2006, Hammell2008, Fesen2011, FM2016}. While S-rich and O-rich ``fast-moving knots" (FMKs) with velocities ranging between 7000 and 15000 km s$^{-1}$ have been identified in Cas~A's northeast and southwest jet regions, several dozen N-rich fast-moving knots have been detected out beyond the remnant's bright ejecta ring and X-ray detected forward shock front with nearly isotropic ejection velocities of $8000-10500$ km s$^{-1}$ \citep{Fesen2001,Fesen2006}. The detection of these high-velocity knots tends to be limited to those ones moving nearly in the plane of the sky (with a max offset of 30\,degrees), which maximises the ability to detect the shock-excitation and ionisation of these ejecta knots when they encounter circum- or interstellar material \citep{Fesen2001,Fesen2011}. 

While the distribution of S-rich and O-rich FMKs tends to be clustered in the northeast and southwest jets of Cas\,A, the observed distribution of the near-tangential N-rich FMKs covers a much wider range of position angles, which suggests that these N-rich knots were ejected isotropically, which is also reflected in their nearly uniform space velocities 
($\approx8000-10500$ km s$^{-1}$). This makes these high-velocity N-rich knots perfect candidates to pierce through the circumstellar shell, clearing the material inside the shell and creating mini-shock waves travelling perpendicular to the direction of the ejecta knot's motion.

The average shock velocities $v_{\text{shock}}=-3500$ km s$^{-1}$ --- as inferred from X-ray spectra of the GM \citep{Vink2024} --- indicate that the expansion has been slower on the near side of the remnant, similar to the slow shock velocities measured towards the south (PA of 190 degrees, \citealt{Vink2022}). However, it is unclear whether shock velocities have been consistently lower compared to the other directions. We conservatively estimate that the forward shock position along the line of sight is similar to the location of the forward shock measured in the plane of the sky ($2.8$\,pc, \citealt{Vink2022}). 
N-rich FMKs would have impacted a shell of circumstellar material positioned at a distance of $2.8$\,pc from the explosion center about $\sim274-304$ years post-explosion. That would mean that the impact already occurred around $\sim$45-75 years ago, and that the material in the circular structures has been expanding since. 

These outer N-rich ejecta knots appear quite small on HST images but are in reality hundreds of AUs in size. The average angular size of these fast-moving knots is 0.1$\arcsec$ \citep{Fesen2016}, but there are several HST resolved N knots with sizes as large as 0.3$\arcsec$ (i.e., around 1000 AU). The apparent clustering of GM holes in certain areas is also consistent with the grouping of N-rich ejecta knots around the periphery of Cas~A 
\citep{Fesen2001,Hammell2008}. Combining their size with their fairly high densities ($n_{\text{e}}\sim2-10\times10^{3}$ cm$^{-3}$, \citealt{Fesen2001}), these hypersonic ejecta projectiles should be capable of making large expanding impact rings in the GM. Compared to the observed size of the ring, we estimate that average shock velocities of $\gtrsim$200 km s$^{-1}$ are required to expand the puncture to the size of the present-day circular structures. 

Alternatively, it is possible that compression waves originating from the displacement of the CSM material lead to the increase of density and hence emission measure at the border of the holes. Unless the FMKs arrived at exactly the same time, the roughly similar sizes of the rings in the GM suggest that the ring size is where the expansion stalls given the transverse momentum deposited in the collision. In this case, the ring would be the result of pileup of material rather than shock heating. 

The observed line emission interior to the dust-emitting part of the rings and the predominantly warm dust emission in the GM rings (compared to the remainder of the GM) would suggest that shock heating and dust destruction by shock waves are relevant, favoring the shock wave scenario.
Unfortunately, we are not able to further test these two scenarios by estimating the velocities of the shock waves that created the circular holes since the MRS spectra are dominated by radial gas motions and hardly trace the tangential motion of these shock or compression waves in the plane of the sky. Furthermore, the material in these rings would cease to expand upon forward shock impact, and we can expect these rings to gradually start shrinking. 

The formation of complete or partial ring structures will depend both on the density of the CSM and its density fluctuations. The absence of full ring structures can likely be attributed to (i) intrinsic density variations in circumstellar shells, (ii) an impact angle of the ejecta bullet that is not perpendicular to the CSM layer or (iii) multiple impacts by a tight cluster of fast-moving ejecta knots which have diluted the densities in these regions. The GM location that is pockmarked with a dozen of holes (see position (d2) in Figure\,\ref{FigGMZoom}) is likely an example of a complex region where several fast-moving ejecta knots impacted the CSM, and where the density is lower (either intrinsically or due to the earlier impacts) creating these partial ring structures. 
 
The multi-ring structure (see Position (d3) in Fig.\  \ref{FigGMZoom}) can also be interpreted in view of this scenario. The negligible radial velocity of the QSF associated with this structure ($v_{\text{r}}$=6$\pm$22 km s$^{-1}$ for QSF\,102, \citealt{Koo2018}; see also the right panel of Figure \ref{FigQSFs}) and the substantial tangential velocity component (1300 km s$^{-1}$) inferred from a proper motion study ($\sim$0.08\,$\arcsec$\footnote{The proper motion is measured from the positional shift in the intensity-weighted centers of the knot in the 2008 and 2013 images. The largest source of uncertainty is associated with the change in morphology due to intensity variations within the knots, and is difficult to quantify.}, position angle of -60$^{\circ}$, \citealt{Koo2018}) would imply that the circumstellar material in this region is tilted with respect to the line-of-sight. In this case, the multiple ring structure could reflect different layers of circumstellar material that are punctured by the same fast-moving ejecta knot. The multi-ring structure is only clearly visible in the NIRCam/F162M image, likely originating from [Fe\,{\sc{ii}}]\,1.644\,$\mu$m line emission. The lack of dust continuum emission for most of the partial rings in this region may suggest that dust has been destroyed (which would explain the presence of Fe in the gas phase). The black arrow overlaid on the NIRCam/F162M image of position (d3) in Figure \ref{FigGMZoom} outlines the potential trajectory from a fast moving ejecta knot tracing back to the center of explosion. 

The good agreement between the alignment of the rings in this position and the trajectory of ballistically moving ejecta knots supports the scenario of ejecta bullets impacting on CSM shells. We note that \citet{Braun1987} already witnessed bow shocks in radio maps resulting from ejecta clumps impacting on decelerated shell material. While these bow shocks result from large, more diffuse condensations of ejecta material, they postulate that small, high-density fragments would have little dynamical impact on the shell, and would leave expanding rings of compressed material. 

Since little mid-infrared dust continuum emission appears to be arising from inside the GM rings, this would suggest that only a thin layer of CSM material is emitting. If an ejecta knot ploughs through a layer of CSM material at a given angle with respect to the line-of-sight, we would expect part of the ring to be filled with emission from a CSM layer that is not parallel with the plane of the sky. The absence of such filled rings suggests that the CSM layer must be intrinsically thin ($<0.32"$) --- where we used the FWHM of the highest resolution MIRI/F1000W image in which the rings are unresolved. This would suggest that the dense CSM material has been lost during a brief mass loss episode that lasted $<300$ years if we assume a typical mass loss velocity for RSGs (10 km s$^{-1}$). The radial velocities of QSFs instead suggest the material has been lost during a faster (100 km s$^{-1}$) mass loss episode, which would restrict the mass loss duration to $<30$ years. These values are not unrealistic given the mass loss variability observed in various RSGs (e.g., \citealt{Humphreys2022}). 

Alternatively, the thickness of the observed GM rings could be representative of the dust destruction length scale. This dust destruction length scale would depend on the average shock speed, the gas density and the grain size of the dust. The observations would require the dust to be destroyed within 30 years time after shock passage if we assume that the material in the ring is flowing at speeds of 100 km s$^{-1}$, respectively. To destroy the small grains ($<0.01$\,$\mu$m) that preferentially emit at these mid-infrared wavelengths, we would require dust grains to be exposed to the hot X-ray emitting gas ($>10^{6}$ K), where these small grains can be efficiently sputtered. Such dust destruction timescales for small grains are not unrealistic (e.g., \citealt{Kirchschlager2019,Kirchschlager2023}). Using the formula from \citet{Hu2019} for the sputtering timescale:
\begin{equation}
t_{\text{sput}}~\approx~0.33\text{Myr} \left(\frac{a}{\mu m}\right) \left(\frac{n_{\text{H}}}{cm^{-3}}\right)^{-1}\left(\frac{Y}{10^{-6}~\mu\text{m}~yr^{-1}~cm^{3}}\right)
\end{equation}
we estimate an average destruction timescale of $\sim$33 years for a 5\,nm grain, a density of 50 cm$^{-3}$ (based on the estimates of the X-ray emitting gas from \citealt{Vink2024}) and a total sputtering yield of $Y_{\text{tot}}$=10$^{-6}$ $\mu$m yr$^{-1}$ cm$^{3}$. In this scenario, we pick up emission of dust grains that are rapidly being destroyed (on timescales of several tens of years) in the GM rings after the shock impact. 

\begin{figure*}
\centering
\includegraphics[width=1.0\textwidth]{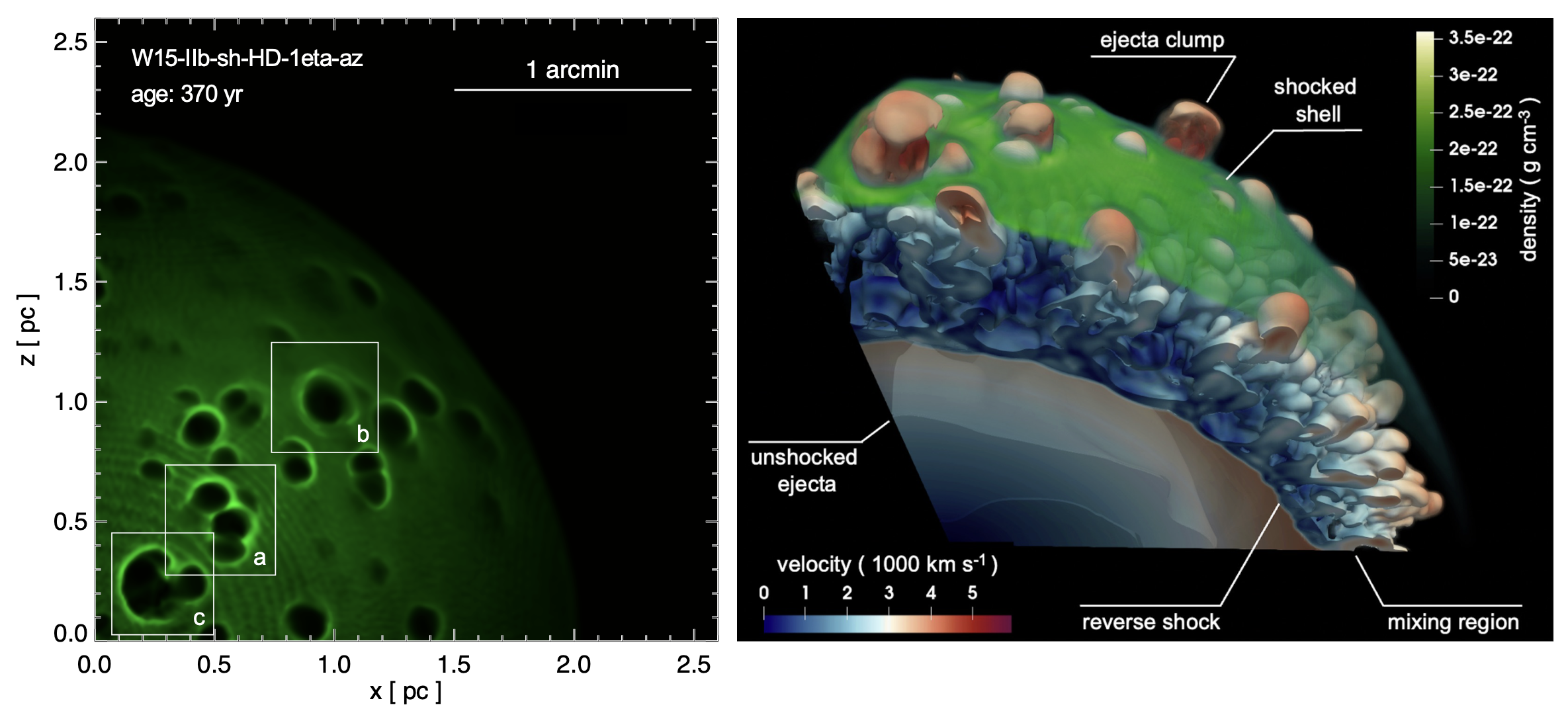}
\caption{Structure of a circumstellar shell after interaction with the remnant at the age of Cas\,A, as derived from a 3D hydrodynamic simulation (run W15-IIb-sh-HD-1eta-az in \citealt{Orlando2022}). Left panel: Three-dimensional volume rendering of the particle density of the shocked shell material (density integrated along the line of sight). Only the upper right quadrant of the numerical domain is shown. The three white squared regions mark holes which are shown enlarged in Figure~\ref{compare_obs_mod} in comparison with similar structures observed with \textit{JWST}. Right panel: Detail of the structure of ejecta and their interaction with the shell at the age of Cas\,A. The semi-opaque irregular isosurfaces correspond to a value of ejecta density, which is at 1\% of the peak density; their colors represent the radial velocity in units of $1000$\,km\,s$^{-1}$ on the isosurface (the color coding is defined at the bottom left corner of the panel). The three-dimensional volume rendering in green color describes the particle density of the shocked shell material (see color bar on the right of the panel).}
\label{FigRingSim}%
\end{figure*}

\subsubsection{\textbf{Scenario IIIb}: Hydrodynamic instabilities create protruding ejecta fingers \textit{after} forward shock impact}
Rather than fast-moving ejecta bullets creating holes in the GM prior to the forward shock impact, it is possible that the forward shock impact triggers hydrodynamic instabilities (Rayleigh-Taylor, Richtmyer-Meshkov and Kelvin-Helmholtz shear instability) at the contact discontinuity that create fingers and clumps of ejecta extending beyond the contact discontinuity to the shocked dense shell of CSM. These fingers protrude into the shocked shell material, which produces holes in the shell similar to the ones observed in our \textit{JWST} observations. 

3D hydrodynamic simulations \citep{Orlando2022} were performed prior to the discovery of these holes in the circumstellar GM structure with \textit{JWST} and were designed to model Cas\,A's remnant evolution from the collapse of the stellar core to the actual age of Cas\,A (see also \citealt{Wong2017, Orlando2021}), including the interaction of the supernova blastwave with a massive circumstellar shell. 

\begin{figure}
\centering
\includegraphics[width=0.45\textwidth]{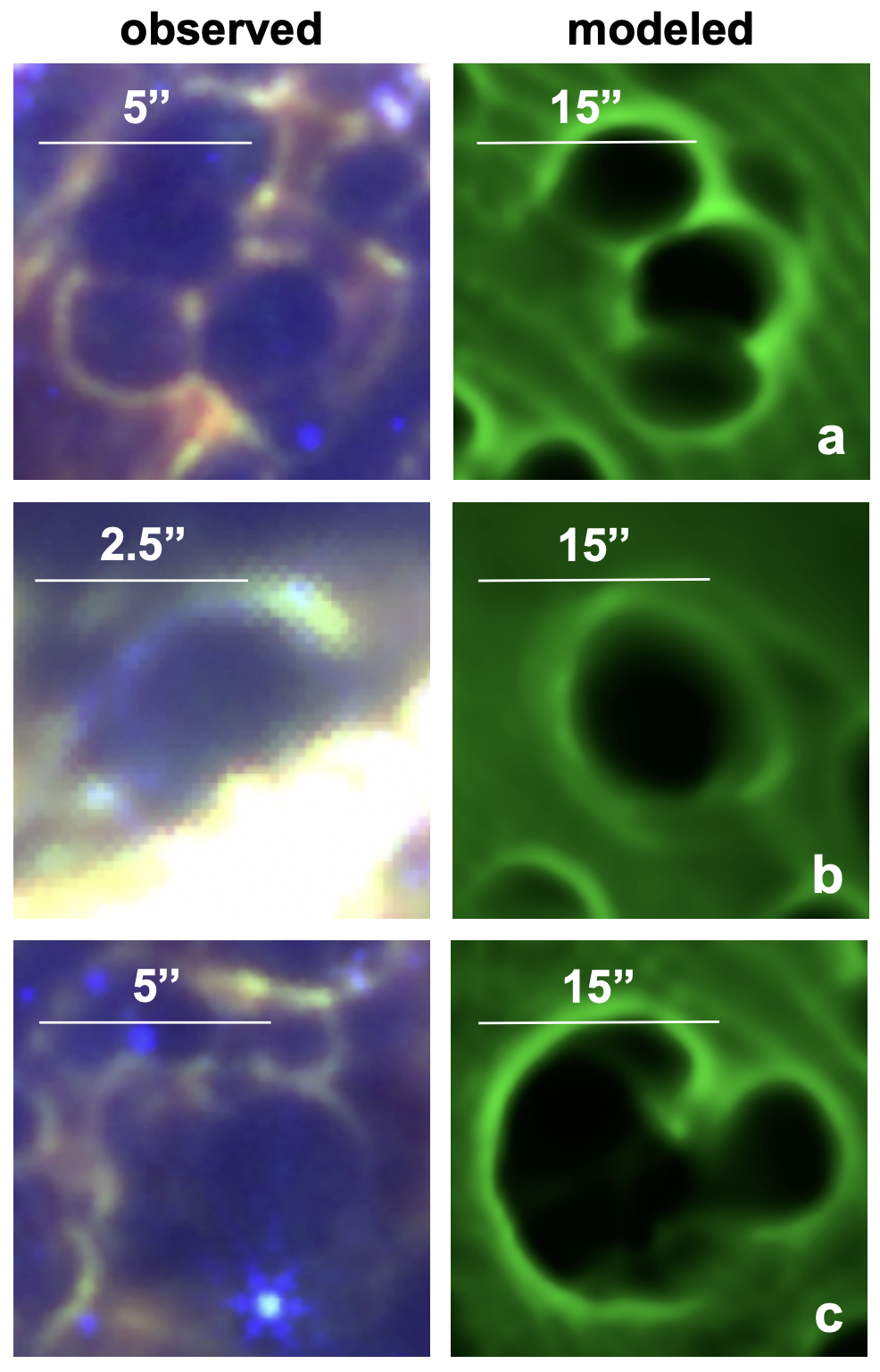}
\caption{A comparison between holes and rings observed with \textit{JWST} (left panels) and analogous structures predicted by hydrodynamic simulations (right panels; \citealt{Orlando2022}) and identified in the white squared regions in Figure~\ref{FigRingSim}.} 
\label{compare_obs_mod}%
\end{figure}

According to these simulations, the asymmetric circumstellar mass loss resulted from a massive eruption in the progenitor star roughly 10,000 to 100,000 years prior to the explosion, and the remnant first encountered the shell roughly 200 years after the explosion. According to the model, the densest portion of the shell is located on the blue-shifted near side of the remnant, consistent with the position inferred for the GM. The ensuing shock accelerated the circumstellar shell material to velocities  of around 
$3000$~km~s$^{-1}$, in agreement with the values inferred from the analysis of X-ray observations \citep{Vink2024}. Soon after it was shocked, the shell started to interact with filamentary structures and clumps of ejecta, rapidly expanding from the contact discontinuity towards the forward shock. These protruding fingers and clumps penetrated the shocked shell material, forming circular holes within its structure. Figure \ref{FigRingSim} shows a detailed view of the shocked shell's structure (left panel) at the age of Cas\,A, derived from model W15-IIb-sh-HD-1eta-az in \cite{Orlando2022}, along with the mechanism driving the formation of the holes (right panel). 

The material within the voids is displaced by the advancing fingers and clumps, gradually accumulating along the hole's periphery, resulting in higher density in those regions (see left panel of Fig.\ ~\ref{FigRingSim}). The cavities and rings formed in this way display shapes remarkably similar to those witnessed with \textit{JWST} (see the enlarged regions in Fig.\ ~\ref{compare_obs_mod}). More specifically, we observe that the model can naturally reproduce, in addition to full regular ring structures, regions pockmarked with several holes partially overlapping each other (upper panels in Fig.\ ~\ref{compare_obs_mod}), as observed in region (d2) in Figure~\ref{FigGMZoom}, multi-ring structures (middle panels in Figure~\ref{compare_obs_mod}) as observed in region (d3) in Figure~\ref{FigGMZoom}, and partial rings (lower panels in Figure~\ref{compare_obs_mod}), as observed, for instance, in Figure~\ref{FigZoomRing}. These features originate from ejecta fingers and clumps that extend from the contact discontinuity and interact with the shocked shell, even though the modeled shell is not characterized by small-scale density fluctuations.

Model simulations also reveal that, on average, the radial velocity of the ejecta structures that interacted with the shell is lower than in regions where such interaction did not occur, due to the passage of the reflected shock from the shell that decelerated the ejecta clumps. The simulated holes have diameter sizes in the range of $\approx 3''-18''$ ($\sim 0.05-0.3~$pc), significantly  larger than those observed in the GM (see Figure~\ref{compare_obs_mod}). This difference in size primarily stems from the limited spatial resolution of current simulations, which impedes an accurate description of the small-scale structure of the ejecta. To accurately capture the size of the observed holes, dedicated simulations with higher spatial resolution will be necessary (Orlando et al.\ in preparation).\\
\newline
With the present data, we are unable to select a preferred scenario. One of the main differences between scenarios IIIa and IIIb is the creation mechanism for the GM rings and the associated timescale. If we assume that the holes were created prior to interaction with the supernova blastwave by fast-moving ejecta bullets that move ahead of the forward shock (Scenario IIIa), this would create a time gap of $\sim$$15-45$ years between the creation of the holes and the forward shock impacting on this circumstellar material. While there is evidence for the existence of such FMKs, it is unclear whether the forward shock interaction would be able to maintain the morphology of the circular rings. There is evidence the GM has been shocked already for at least 30 years, which would require the circular rings to have incredibly high densities in order to avoid distortion of these circular structures upon shock impact.

This problem could be remedied if we are probing a thin layer of recently shocked material in which small grains are not yet completely destroyed. On the other hand, we would expect to observe ring structures which are partially filled when observing fingers of ejecta clumps penetrating through a shocked circumstellar layer inclined with respect to the line-of-sight. In the simulations, the shell has a thickness of $\sim$0.02\,pc, whereas the observations would constrain the shell thickness to be 10 times smaller. It is possible that the destruction timescale also has an impact here on what we are observing. It is furthermore unclear whether the evolution of the hydrodynamical instabilities --- resulting in the creation of GM-like ring structures --- could have been influenced by the 3D nature of the progenitor and its 3D mass loss instead of the assumed one-dimensional progenitor model \citep{Woosley1995}, whose core-collapse was evolved through several simulation stages \citep{Wong2017} that finally led to the input model of \citet{Orlando2021,Orlando2022}. Future work will need to study the impact of some simplifications in the progenitor and SN explosion models (e.g., non-turbulent CSM/ISM structure and smooth shells) on the late-time interaction with the reverse shock and CSM, and the subsequent formation of ring-like structures similar to the observations with JWST in the GM. Currently, ongoing efforts (Orlando et al. in prep.) are dedicated to study the combined effects of efficient radiative cooling (which makes ejecta fingers and clumps thinner and denser) and the magnetic field's confinement of ejecta structures (which enhances radiative losses) on the rings' shapes and sizes. 

\section{Discussion}
\label{CSM.sec}

\subsection{Cas A's mass loss history}
In this paper and a companion paper \citep{Vink2024}, we argue that the newly detected GM structure pockmarked with holes corresponds to circumstellar dusty material that was lost by the progenitor star about 3$\times10^{4}$ to 10$^{5}$ yrs prior to explosion. A valid question to ask ourselves is how the GM is related to other mass loss phases. 

Comparison of dust SEDs (De Looze et al.\,in prep.) and X-ray spectra \citep{Vink2024} suggests that the GM closely resembles the properties of the southern arc of circumstellar material that is currently impacted by the supernova blast wave and that was already identified from early optical observations \citep{vandenBergh1983,vandenBergh1985,Lawrence1995}. A significant number of QSFs is located in the southern circumstellar arc or the GM, which may suggest that QSFs are linked to the same mass loss phase that ejected this circumstellar material. In addition, several QSFs are located along the shocked main ejecta shell in the north or scattered randomly within the main ejecta shell. We suggest these QSFs may be connected to circumstellar material that is interacting with high-velocity ejecta knots -- which are prominently seen in the main ejecta shell in the north \citep{Fesen2011}. 

In Figure \ref{FigQSFs} (left panel), we show that the GM's location fits in well within the elongated shell-like structure of dense shocked CSM knots discerned through their H$\alpha$ emission \citep{Fesen2001}. This picture gives high credibility to a scenario where the GM and QSFs were ejected by the Cas~A progenitor star during the same mass loss episode. This suggests that the mass loss experienced by the progenitor star resulted in one-sided mass loss on the near side of the Cas\,A system. In this view, QSFs most likely represent small clumps of especially high densities in the remnant's circumstellar medium. 

While the GM appears to resemble the properties of the circumstellar dust in the southern arc, its green color clearly distinguishes the GM from the yellow, orange and red colored circumstellar clouds located in the north, east and west of Cas\,A. In a follow-up work (De Looze et al.\,in prep.), we will discuss how the grain composition and size varies for circumstellar dust formed from material lost during distinct mass loss episodes.  
The diffuse red circumstellar clouds in the north, east and west of Cas\,A also stand out for not hosting any QSFs \citep{Koo2018}. A close examination of the NIRCam/F162M image confirms that --- apart from one region that shows evidence for interaction with ejecta knots --- there is no trace of Fe emission in these red circumstellar clouds (Koo et al.\,in prep.). 

While the bulk grain material appears to be similar, the dissimilar \textit{JWST} colors and dust characteristics (e.g., grain size, relative grain abundance) supports the distinct nature of these circumstellar clouds (De Looze et al.\,in prep.). The absence of any detected QSFs -- as traced by NIR [Fe\,{\sc{ii}}] 1.644\,$\mu$m line emission \citep{Koo2023} or optical H\,$\alpha$ and [N\,{\sc{ii}}] $\lambda\lambda$6548,6583 line emission \citep{vandenBergh1985}
-- suggests that the circumstellar material towards the north, east and west of Cas\,A is less dense, as already suggested by \citet{Koo2023}. Analyses of the X-ray spectra also suggest that the GM has elevated densities compared to the circumstellar dust in the north \citep{Vink2024}. 

\subsection{Highly asymmetric mass loss?}

Based on the accompanying X-ray analysis \citep{Vink2024}, we argue that the QSFs and GM are located in front of Cas\,A. With no evidence for a counterpart behind Cas\,A -- based on the lack of similar interactions on the rear side, this would require highly asymmetric mass loss from the progenitor star. This argument is further supported by hydrodynamic simulations that show evidence of interaction of Cas\,A with an asymmetric dense circumstellar shell on the blueshifted nearside of the remnant (\citealt{Orlando2022}). 
There is little evidence of single stars experiencing significant asymmetric mass loss (see examples in \citealt{2001AJ....121.1111S,Montarges2021,Humphreys2022}). Hence, a binary scenario may be required to account for the highly asymmetric one-sided mass loss that would have produced the observed distribution of the GM and the QSFs (see also \citealt{Koo2023}). 
The one-sided mass loss may point to specific mass loss scenarios, like eccentric orbits or sudden mass loss instead of smooth Roche Lobe overflow in a binary system (e.g., \citealt{Lau2022,2024MNRAS.531.3391L}). However, it is important to note that the asymmetric mass loss observed in some AGB stars and RSGs \citep{Cox2012} has been attributed to alternative explanations, including large convective cells \citep{Lim1998}. 
Since deep searches have failed to detect a surviving companion star for Cas~A, and this has been interpreted as evidence that the binary progenitor system of Cas~A merged prior to the supernova explosion \citep{Kochanek2018,Kerzendorf2019}, there may be a connection between this merger phase and the extremely dense mass loss phase associated with the GM and the QSFs. A similar merging scenario has been suggested for SN~1987A's pre-explosion stellar evolution \citep{Morris2007}. However, the mass loss in SN~1987A's merging progenitor system resulted in a more or less axially-symmetric mass distribution, not one-sided akin to the mass loss observed in Cas~A's progenitor system. In planetary nebulae, highly asymmetrical mass loss with one side much denser than the other is attributed to triple-star interaction (e.g., \citealt{Bear2017}). Future observations and numerical modelling will be crucial to study the complex nature of the Cas A progenitor system and to unravel the origin of the highly asymmetric mass loss phase that resulted in the formation of the QSFs and GM. 

\section{Conclusions}
\label{Concl.sec}
Extensive mapping of Cas\,A with \textit{JWST} reveals a unique structure seen in projection toward the center that is pockmarked with dozens of $1''-3\arcsec$ sized circular holes surrounded by rings of enhanced emission \citep{Milisavljevic2024}. The green color in multi-band \textit{JWST} imaging (originating from MIRI/F1130W and F1280W emission) makes this ``Green Monster" (GM) stand out from the remnant's other emission components. We present a multi-wavelength view -- including \textit{JWST} NIRCam (F162M, F356W, F444W) and MIRI imaging of the entire GM structure, and NIRSpec and MRS spectroscopic observations of a specific ring region, which allows an in-depth study of the origin of the GM emission. 

We find that the GM corresponds to dusty circumstellar material located on the near-side of Cas\,A that has been impacted by the forward shock. The detection of emission lines such as Br\,$\alpha$ 4.05\,$\mu$m in the NIRSpec spectra and [Ne\,{\sc{ii}}] 12.81\,$\mu$m, [Ne\,{\sc{iii}}] 15.56\,$\mu$m, and [Fe\,{\sc{ii}}] 17.94\,$\mu$m in the MIRI MRS spectra coinciding with part of the ring structure are consistent with a circumstellar origin. The low radial velocities inferred for these lines -- ranging between $-50$ and 0 km s$^{-1}$ -- provide additional evidence that we are tracing circumstellar material. The distribution of dense circumstellar clumps -- better known as quasi-stationary flocculi -- are in general alignment  with the GM's location.

The analysis of X-ray data in a companion paper \citep{Vink2024} further corraborates this scenario, finding similarity between the X-ray spectra in the GM and in previously identified shocked circumstellar dust clumps south of Cas\,A. The blueshift of the X-ray emission furthermore suggests the interaction with the supernova blastwave happens on the near side of the remnant, which is consistent with the domination of negative radial velocities of the QSFs \citep{Koo2018}. 

The most striking and unexpected features of the GM are the numerous small circular holes that are distributed unevenly across the circumstellar structure. We believe the holes were created upon interaction of ejecta with circumstellar material seen as the GM feature in colored JWST images. However, the exact timing for these interactions remains unclear. The impact of fast-moving ejecta knots could have taken place several tens of years prior to the forward shock impact. Upon impact, these knots would instigate shock- or compression waves in the perpendicular direction from the ejecta knots' motions with circumstellar material piling up and creating the (partial) ring structures. The obvious candidate capable of piercing through dense layers of circumstellar dust are Cas~A's N-rich ejecta knots that move with fairly uniform space velocities of 
8000--10500 km s$^{-1}$ and appear to have been ejected nearly isotropically from the explosion center \citep{Fesen2001, Hammell2008}. 

Alternatively, protrusions in the remnant's expanding ejecta can create holes in circumstellar shells soon after the impact by the forward shock due to ejecta ``fingers'' created through hydrodynamical instabilities that extend from the ejecta through the contact discontinuity to the shocked circumstellar material, as demonstrated in recent numerical simulations \citep{Orlando2022}. Future observations and simulation efforts are needed to constrain the exact mechanism and formation timescale for these holes. 

The detection and characterisation of the GM structure with \textit{JWST} provides further evidence for a highly asymmetric mass loss episode in Cas\,A's progenitor star shortly ($3\times10^{4}$ to 10$^{5}$ yrs) before the explosion. This conclusion is in line with previously identified shocked dense circumstellar QSFs \citep{Koo2023}. The GM and QSFs share several characteristics, which may suggest that these dense CSM structures also share a common same mass loss phase experienced by Cas~A's progenitor star. Future \textit{JWST}/NIRSpec MOS spectroscopic observations for a variety of GM and QSF regions will be crucial to gain insights into the velocity structure, the physical properties and the elemental abundances of this new CSM component to understand the late evolutionary phases of the Cas~A progenitor mass-loss prior to its explosion.

\begin{acknowledgments}
This work is based on observations with the NASA/ESA/CSA James Webb Space Telescope obtained at the Space Telescope Science Institute, which is operated by the Association of Universities for Research in Astronomy, Incorporated, under NASA contract NAS5-03127.  Support for program number JWST-GO-01947 was provided through a grant from the STScI under NASA contract NAS5-03127. The data presented in this article were obtained from the Mikulski Archive for Space Telescopes (MAST) at the Space Telescope Science Institute. The specific observations analyzed can be accessed via \dataset[doi: 10.17909/szf2-bg42]{https://doi.org/10.17909/szf2-bg42}.
I.D.L. and J.C. acknowledge funding from the Belgian Science Policy Office (BELSPO) through the PRODEX project “JWST/MIRI Scence exploitation” (C4000142239). I.D.L., F.K., N.S.S. and T.S have received funding from the European Research Council (ERC) under the European Union’s Horizon 2020 research and innovation programme DustOrigin (ERC-2019-StG-851622). N.S.S acknowledges the support from the Flemish Fund for Scientific Research (FWO-Vlaanderen) in the form of a post-doctoral fellowship (1290123N). 
D.M.\ acknowledges NSF support from grants PHY-2209451 and AST-2206532.
Work by R.G.A. was supported by NASA under award No.
80GSFC21M0002.
T.T. acknowledges support from NSF grant 2205314 and JWST grant JWST-GO-01947.031.
B,-C.K. acknowledges support from the Basic Science Research Program through the NRF funded by the Ministry of Science, ICT and Future Planning (RS-2023-00277370). J.R. acknowledges funding from JWST-GO-01947.032 and the NASA ADAP grant (80NSSC23K0749).
J.M.L. was supported by JWST grant JWST-GO-01947.023 and by basic research funds of the Office of Naval Research. S.O. acknowledges support from PRIN MUR 2022 (20224MNC5A) funded by European Union – Next Generation EU.
H.-T.J. is grateful for support by the German Research Foundation (DFG) through the Collaborative 
Research Centre ``Neutrinos and Dark Matter in Astro- and Particle Physics (NDM),'' Grant No.\ SFB-1258-283604770, and under Germany's Excellence Strategy through the Cluster of Excellence ORIGINS EXC-2094-390783311.
M.M. acknowledges support from the STFC Consolidated grant (ST/W000830/1).
\end{acknowledgments}







\appendix

\section{Spectral decomposition}
\subsection{Line profile analysis}
\label{LineProfile.sec}
We present an overview of the spectral decomposition of the NIRSpec and MIRI/MRS spectra in different velocity components for each of the bright lines (see Figure \ref{FigLineProfiles}) detected in the MIRI MRS pointing targeting a bright ring in the GM. To limit the contamination by other velocity components, we extracted the emission from a $2.5\arcsec \times 2.5\arcsec$ box centered on the ring. The \textit{ejecta-dominated} lines ([S\,{\sc{iv}}] 10.51\,$\mu$m, [S\,{\sc{iii}}] 18.71\,$\mu$m, [O\,{\sc{iv}}] 25.89\,$\mu$m) show several velocity components ranging between -5000 km s$^{-1}$ and +5000 km s$^{-1}$. For sulfur, the brightest lines are blueshifted with radial velocities ranging between -2500 and 0 km s$^{-1}$. There is a good correspondence between the velocity components detected in both sulfur lines. The predominance of blueshifted components may reflect asymmetries in the explosion and/or could be the consequence of shock interactions on the near side of Cas\,A due to a dense CSM that has created a reverse shock early on that is currently interacting with the expanding ejecta. For the redshifted velocity components, there are peaks around +950, +1550 and +4000 and +4400 km s$^{-1}$ for the sulfur and oxygen lines. The oxygen line profile shows a very prominent line around $\sim$+5000 km s$^{-1}$ without any associated sulfur emission, but does show [Ne\,{\sc{ii}}], [Ne\,{\sc{iii}}] and [Ne\,{\sc{v}}] line emission at those velocities that is spatially coincident with a high-velocity ejecta filament that runs up against the base of the ring structure (see Figure \ref{FigLineMapsFil}). The absence of any obvious sulfur line emission corresponding with this high-velocity ejecta filament may suggest that it originates from the O/Ne layer of the progenitor star, whereas the other ejecta filaments instead originate from layers where there has been mixing between the O/Ne and Si/S layers. Apart from this high velocity ejecta filament, there is excellent correspondence between the S and O line velocity components, which leaves little room for a contribution from the [Fe\,{\sc{ii}}]\,25.99\,$\mu$m line. We stress that it is unlikely for the [Fe\,{\sc{ii}}]\,25.99\,$\mu$m line to contribute significantly to the v=5000 km s$^{-1}$ velocity component due to the absence of any other iron line emission ([Fe\,{\sc{ii}}]\,17.93\,$\mu$m, [Fe\,{\sc{iii}}]\,22.93\,$\mu$m) at those high velocities within the MIRI MRS wavelength coverage. 

Although the MIRI MRS cubes have been background subtracted, there is preponderance of near rest-frame H$_{2}$ line emission across the entire MIRI MRS field of view that seem unrelated to any of the ejecta and/or circumstellar velocity components. These H$_{2}$ lines map out the warm interstellar medium on the line of sight to Cas\,A. The detection of warm H$_{2}$ emission requires shock excitation or UV pumping for the material to reach temperatures of several 100\,K.

\begin{figure*}
\centering
\includegraphics[width=1.0\textwidth]{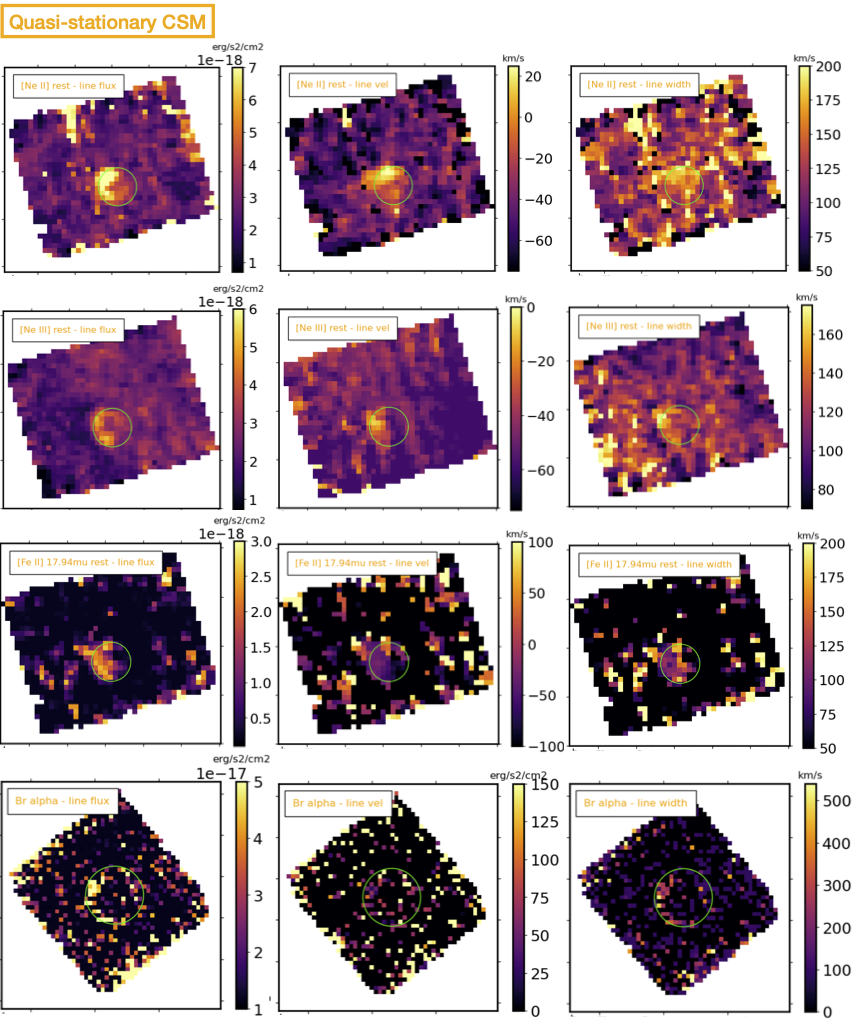}
\caption{Spectral mapping of the GM ring region in position P2 (see Figure\,\ref{FigWebbJudy}) near rest-frame velocities. Maps are shown of the line flux (left column), line velocity (middle column) and FWHM line width (right column) for low-velocity components of the [Ne\,{\sc{ii}}]\,12.81\,$\mu$m (first row), [Ne\,{\sc{iii}}]\,15.56\,$\mu$m (second row), [Fe\,{\sc{ii}}]\,17.94\,$\mu$m (third row) and Br\,$\alpha$ (fourth row) lines. The green circle roughly outlines the position of the ring in MIRI dust continuum maps, with the same circle overlaid on the GM images shown in Figure \ref{FigZoomRing}.}
\label{FigLineMapsRing}%
\end{figure*}

\begin{figure*}
\centering
\includegraphics[width=1.0\textwidth]{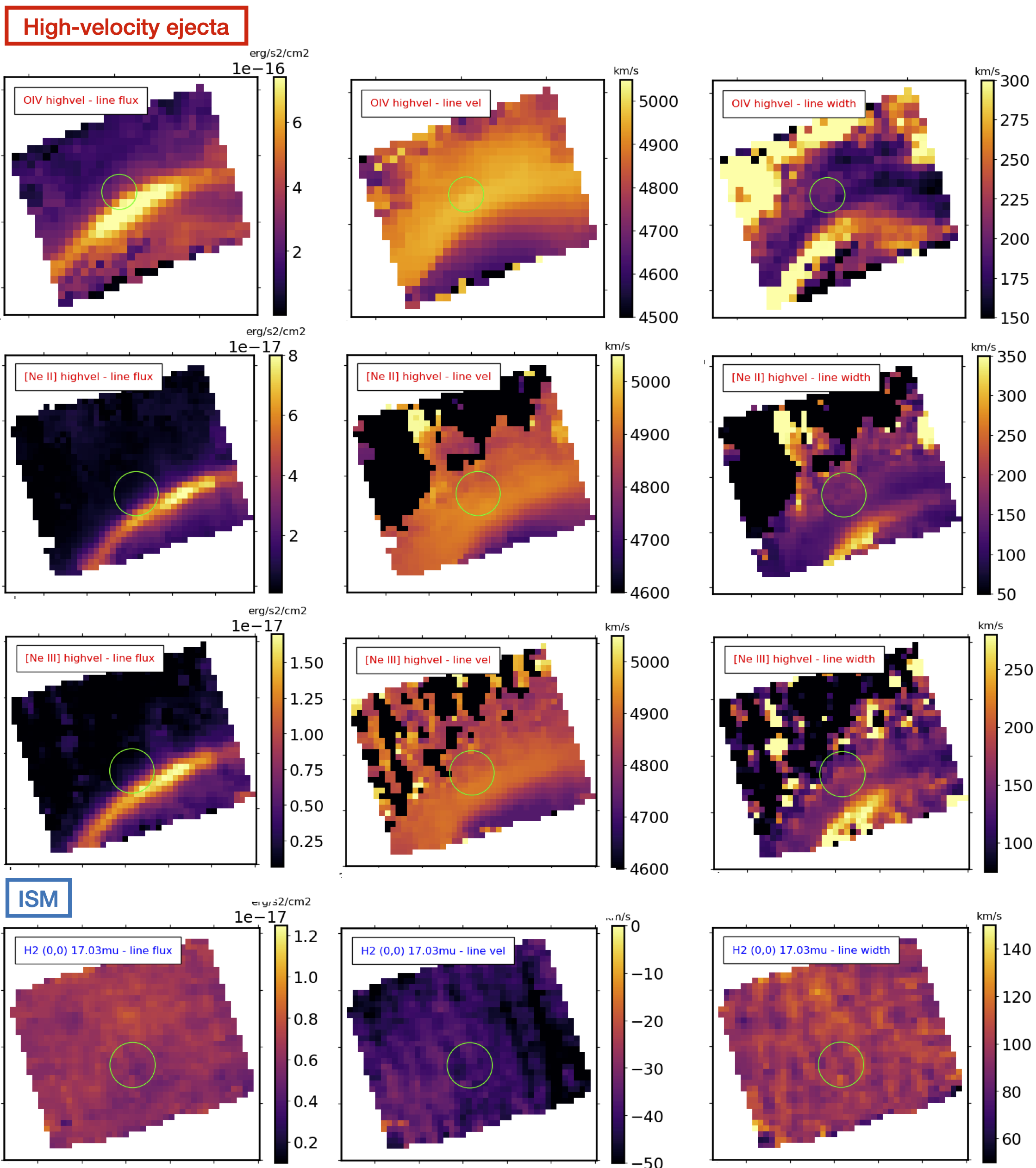}
\caption{Spectral mapping of the GM ring region in position P2 (see Figure\,\ref{FigWebbJudy}) for the high-velocity (v $=5000$ km s$^{-1}$) ejecta component (top panels) and the stationary ISM component (bottom row). For the high-velocity ejecta component, maps are shown of the line flux (top column), line velocity (middle column) and FWHM line width (bottom column) for the [O\,{\sc{iv}}]\,25.89\,$\mu$m (first row), [Ne\,{\sc{ii}}]\,12.81\,$\mu$m (second row), and [Ne\,{\sc{iii}}]\,15.56\,$\mu$m (third row) lines. The line flux, line velocity and line width maps of the H$_{2}$ (0,0) 17.03\,$\mu$m line of interstellar origin are shown in the bottom panels. The green circle roughly outlines the position of the ring in MIRI dust continuum maps (see Figure  \ref{FigZoomRing}) and is overlaid on the line maps to guide the eye. } 
\label{FigLineMapsFil}%
\end{figure*}

Another near rest-frame velocity component traces \textit{quasi-stationary circumstellar material} emitting brightly in several Ne ([Ne\,{\sc{ii}}] 12.81\,$\mu$m and [Ne\,{\sc{iii}}] 15.56\,$\mu$m), Fe ([Fe\,{\sc{ii}}] 1.644\,$\mu$m, and [Fe\,{\sc{ii}}] 17.93\,$\mu$m)  and hydrogen recombination (Br\,$\alpha$ 4.05\,$\mu$m\footnote{Note that other rings in the GM also emit in H\,$\alpha$ (see Figure \ref{FigGMZoom}).}) lines. There is a tentative detection of [Ne\,{\sc{v}}] 14.32\,$\mu$m line emission (see Figure \ref{FigLineProfiles}) near rest-frame wavelengths, but the offset in the line emission's spatial distribution and velocity (-220 km s$^{-1}$) warrants further investigation into the origin of this faint [Ne\,{\sc{v}}] line emission. 
This near-zero km s$^{-1}$ velocity component originates from a partial ring that closely resembles the morphology of the ring visible in the dust continuum maps (see Figure \ref{FigLineMapsRing}). 

\subsection{Line modelling}
\label{LineFit.sec}
Due to the complexity of the \textit{JWST}/MIRI MRS spectra with a lot of material along the central lines of sight in Cas\,A, we fit a number of Gaussian profiles to disentangle the emission of different velocity components. We start by defining line-free wavelength ranges on either side of the line emission and fit a first-order polynomial to subtract the underlying continuum emission. We fit a minimum number of Gaussian components (i.e., one for each velocity component). The best-fit parameters for each Gaussian profile are determined based on a $\chi^{2}$ minimisation routine. The line flux in each spaxel is calculated as the flux integrated over the best-fit Gaussian profile. The radial velocity is calculated based on the Doppler velocity shift of the line center with respect to the rest-frame wavelength in vacuum, whereas the FWHM line width is calculated from the best-fit Gaussian width parameter $\sigma$. To calculate the upper limits, we measure the 1$\sigma$ standard deviation in a line-free region of the spectrum, and calculate the upper limit as the flux integrated over a Gaussian profile with width similar to the other detected lines in that wavelength range and with amplitude equal to 3$\times$ the standard deviation $\sigma$. We followed this procedure for all lines associated with the quasi-stationary (near zero velocity) CSM component that we identified as the GM. Table \ref{Line_fluxes} summarizes the measured line fluxes, heliocentric line velocities and FWHM line widths. Note that we calculated upper limits for the [S\,{\sc{iv}}], [S\,{\sc{iii}}], [O\,{\sc{iv}}] line fluxes, even though there is emission detected around 0 km s$^{-1}$, since this emission seems unrelated to the GM ring detected in dust continuum. 

The [Fe\,{\sc{ii}}]\,1.644\,$\mu$m line flux is inferred from the NIRCam/F162M image using a rather complex procedure. All NIRCam images were first background subtracted. We then isolated the ring extracting the emission within an annulus with inner and outer radius of 0.4 and 0.6\arcsec. We estimated the general background emission from a diffuse synchrotron background using the emission inside the ring. After subtracting this diffuse synchrotron background component, the residual emission will have contributions from synchrotron radiation (as detected in the ring structure in the NIRCam/F356W and F444W images) and from the [Fe\,{\sc{ii}}]\,1.644\,$\mu$m line emission. To calculate the contribution from synchrotron radiation within the ring, we measured the fluxes for the GM ring in the NIRCam/F356W and F444W images within the same annulus (after subtracting a diffuse synchrotron background) and fitted a power-law spectrum with $\alpha=-0.60$ (i.e., the mean value from \citealt{Domcek2021} for this region) to these fluxes. Convolving this power-law spectrum with the JWST/NIRCam filters, we estimate the contribution from synchrotron emission within the ring, which is subtracted to obtain the residual [Fe\,{\sc{ii}}] line emission. Using the F356W flux to estimate the synchrotron contribution in the F162M filter, we are left with a flux of $1.25\times10^{-16}$ erg s$^{-1}$ cm$^{-2}$ before applying any dust extinction correction. Extrapolation from the F444W flux for the ring results in a high predicted synchrotron flux in the F162M filter, making it debatable whether or not there is any residual Fe line emission originating from the NIRCam/F162M image. It is likely that other contributions to the F444W emission (e.g. hot dust emission) may cause these high synchrotron flux estimates. Given the many uncertain factors (i.e., the normalisation and power-law spectrum of the diffuse background, local synchrotron components and dust extinction corrections), and the unavailability of NIRSpec data covering the [Fe\,{\sc{ii}}]\,1.644\,$\mu$m line, we consider the inferred [Fe\,{\sc{ii}}]\,1.644\,$\mu$m flux using the extrapolation of the F356W flux to estimate the synchrotron contribution as an upper limit. We require short-wavelength NIRSpec data in future JWST Cycles in order to gauge the contributions of synchrotron emission and other [Fe\,{\sc{ii}}] lines (i.e., 1.599, 1.664, and 1.677\,$\mu$m lines) that could also make substantial contributions to the F162M filter band \citep{Koo2016}. 

Finally, we need to consider dust extinction corrections for some of the lines. Due to the dense column of interstellar material towards the line-of-sight to Cas\,A, the visual extinction ranges between $A_{\text{V}}$ of 7.5 and 15\,mag \citep{DeLooze2017} depending on the specific sightline. Following the methodology of \citet{Koo2018}, we estimate a visual extinction of $A_{\text{V}}$ = 9.36\,mag from the column density map of \citet{Hwang2012} and assuming $N_{\text{H}}$/$A_{\text{V}}$ = $1.87\times10^{21}$ cm$^{-2}$ mag$^{-1}$. This translates into $A_{\text{1.644}}$=1.73\,mag assuming $A_{\text{1.644}}$/$A_{\text{V}}$ = 1/5.4 \citep{Draine2003}, suggesting that we need to correct our [Fe\,{\sc{ii}}] 1.644\,$\mu$m flux by a factor of 5. In a similar way, the Br\,$\alpha$ emission is corrected for dust extinction assuming $A_{\text{4.05}}=$0.044\,mag.

\begin{table*}
    \centering
    \caption{Near rest-frame velocity line fluxes, heliocentric radial velocities and full-width half maxima (FWHM) of the emission lines showing resemblance with the GM ring targeted with the NIRSpec and MRS observations in P2 
    (see Figure~\ref{FigWebbJudy}). In case of a non-detection, we quote 3$\sigma$ upper limits. Note that the quoted line fluxes have not yet been corrected for dust extinction. The rest-frame wavelength in vacuum was adapted from the \textit{JWST} line list (\url{http://hebergement.u-psud.fr/edartois/jwst\_line\_list.html}).}
    \label{Line_fluxes}    
    \begin{tabular}{ccccc}
    \hline 
         Line & Rest-frame wavelength [$\mu$m] & Line intensity [10$^{-16}$ erg/s/cm$^{2}$] & Line velocity [km s$^{-1}$] & FWHM line width [km s$^{-1}$] \\
         \hline 
         \,[Fe\,{\sc{ii}}] & 1.644 & $\le$1.25 & - & - \\
         Br\,$\alpha$ & 4.05 & 0.29$\pm$0.04 & -19 & 226 \\
         \hline 
         \,[Fe\,{\sc{ii}}] & 5.34017 & $\le$8.3 & - & - \\
         \,[Ne\,{\sc{ii}}] & 12.81355 & 5.0$\pm$0.2 & -24 & 132 \\
         \,[Ne\,{\sc{iii}}] & 15.55505 & 4.4$\pm$0.3 & -44 & 125 \\
         \,[Fe\,{\sc{ii}}] & 17.93603 & 0.8$\pm$0.2 & -42 & 84   \\
         \,[Fe\,{\sc{iii}}] & 22.9250 & $\le$3.0 & - &  - \\
         \,[O\,{\sc{iv}}] & 25.8903 & $\le$5.2 & - & -  \\
         \hline         
    \end{tabular}
\end{table*}

\bibliography{sample631}{}
\bibliographystyle{aasjournal}



\end{document}